\documentclass[12pt,a4paper]{article}
\usepackage[utf8]{inputenc}
\usepackage[T1]{fontenc}
\usepackage{lmodern}
\usepackage[english]{babel}
\usepackage[a4paper,margin=2.5cm]{geometry}
\usepackage{amsmath, amssymb, amsfonts, mathrsfs, bm}
\usepackage{braket}
\usepackage{graphicx}
\usepackage{xurl}
\usepackage{hyperref}
\hypersetup{hidelinks}
\newcommand{\dd}{\mathrm{d}}
\newcommand{\ii}{\mathrm{i}}
\title{Static and dynamical Casimir effects in a wedge within the worldline formulation}
\author{C.~D.~Fosco \\
and \\
B.~C.~Guntsche\\
{\normalsize\itshape Centro At\'omico Bariloche and Instituto Balseiro}\\
{\normalsize\itshape Comisi\'on Nacional de Energ\'\i a At\'omica}\\
{\normalsize\itshape R8402AGP Bariloche, Argentina.}}
\date{}
\begin{document}
\maketitle
\begin{abstract}
We study a real massless scalar field in $d+1$ dimensions confined to a wedge
of opening angle $\alpha$, in the presence of a semitransparent circular arc
that may oscillate radially. Within the worldline formulation, and for an
arbitrary strength $\lambda$ of the $\delta$-potential that models the arc, we
expand the effective action in powers of the radial displacement $R$ of the
arc. In this formulation, all the information about the media is contained in
the propagator of a non-relativistic point particle, so that known results for
that propagator can be used directly. At zeroth order we obtain the Casimir
energy of the wedge, with Dirichlet, Neumann or mixed walls, and the
contribution of the arc in closed form for arbitrary $\lambda$. We show that
the worldline integrand is, point by point, the energy density obtained from
the energy-momentum tensor of the scalar field with curvature coupling
$\xi=1/4$; with this identification, and using dimensional regularization for
the integrated quantities, we recover the known results for the wedge and for
the wedge closed by a Dirichlet arc. At second order in $R$, we reduce the
form factor that determines the imaginary part of the in-out effective action,
and hence the probability of pair creation by the moving arc, to a one-loop
expression involving the radial Green's functions of each angular sector. In
the narrow-wedge limit, each angular sector behaves as a field with a mass set
by the angular quantization, coupled to a moving $\delta$-mirror; for a
Dirichlet arc, the pair creation spectrum of the exterior region approaches
that of a planar massive mirror, while the interior region behaves as a
tapered cavity. Finally, by
decomposing the electromagnetic field into TE and TM scalars, we obtain the
Casimir energy of wedges with perfectly conducting, perfectly permeable and
mixed walls.
\end{abstract}
\section{Introduction}\label{sec:intro}

The Casimir effect has by now been studied in a wide variety of
situations, and a number of exactly solvable geometries are known. Among
them, the wedge occupies a prominent place: it is simple enough to admit
closed-form results, and at the same time it is a building block of more
complicated boundaries (polygonal cavities, cylindrical sectors, edges and
tips of all kinds). In this paper we revisit the wedge, and the wedge closed
by a semitransparent circular arc, from the point of view of the worldline
formulation, both for static and for time-dependent configurations.

At the local level, the vacuum expectation value of the energy-momentum
tensor inside a perfectly reflecting wedge was obtained by Dowker and
Kennedy for a scalar field~\cite{DowkerKennedy}, and by Deutsch and
Candelas for the conformally coupled scalar and for the electromagnetic
field~\cite{DC}. The electromagnetic case was reconsidered in
Ref.~\cite{BrevikLygren}, and local zeta-function evaluations were presented
in Refs.~\cite{NLS_wedge,FP1,FP2}. The energy density of a scalar field with
arbitrary curvature coupling $\xi$, inside a wedge closed by a Dirichlet
circular arc, was obtained in Ref.~\cite{RS}, and the exterior region was
considered in Ref.~\cite{SaharianTarloyan}. These local quantities are not
unique: the energy-momentum tensor admits an improvement term proportional
to $\xi$, and different choices of $\xi$ lead to different energy densities
and, as we shall see, to different radial energy distributions.

At the global level, the Casimir energy of a semicircular cylinder was found
to be divergent~\cite{NLS}, a divergence that was traced back to the
edges~\cite{Dowker_edges}. An unremovable divergence associated with the
corners where the arc meets the walls was also found for the
electromagnetic wedge closed by a cylindrical
arc~\cite{BEM_wedge1,EBM_wedge2}, which was extended to nonzero temperature
in Ref.~\cite{EBM_temperature}; a wedge with semitransparent walls, closed
by a Dirichlet arc, was studied in Ref.~\cite{MWK}. The energies of the full
cylinder, for Dirichlet and Neumann scalar fields and for the
electromagnetic field, are known~\cite{GosdzinskyRomeo}. The relation
between the total energy and the torque on the walls has been the subject
of a discussion on a possible torque anomaly~\cite{FMT,Dowker_torque},
which was examined in annular sectors in
Refs.~\cite{Milton_annular1,Milton_annular2}; the special role of the
coupling $\xi=\frac{1}{4}$, for which the energy density contains no
product of gradients, was emphasized in Ref.~\cite{EFM}.

The worldline formulation~\cite{Gies_WL,Worldline_Review} represents the
effective action in terms of the propagator of a non-relativistic particle
moving in the background defined by the media. It has been used extensively
in numerical evaluations of Casimir energies~\cite{Gies_WL} and, more
recently, to study the dynamical Casimir effect for planar media of
arbitrary strength~\cite{FG_WL}. The latter complements functional
approaches to the dissipative effects of moving perfect and imperfect
mirrors~\cite{FLM_2007,FLM_2011,FG_Dirichlet,FG_Neumann,FG_Amplitudes}. In
this context, the wedge closed by an arc is a natural next step beyond the
plane: the angular sector is exactly solvable at the level of propagators,
and the arc, modeled by a $\delta$-potential of arbitrary strength
$\lambda$, can undergo radial oscillations.

Our aims in this work are the following. First, we show that the whole
static problem, namely, the energies of wedges with Dirichlet, Neumann or
mixed walls, with and without the semitransparent arc, follows from the same
worldline propagators, with the arc contribution obtained in closed form for
arbitrary $\lambda$ (Sect.~\ref{sec:zero}); for perfectly reflecting arcs,
the only divergence left is a logarithmic one, associated with the corners
and independent of $\alpha$.
Second, we show that the worldline integrand is, point by point in the
interior of the wedge, the local energy density of a scalar field with
curvature coupling $\xi=\frac{1}{4}$; this identification turns the known
arbitrary-$\xi$ results of Refs.~\cite{DC,RS} into exact evaluations of the
worldline integrand (Sect.~\ref{sec:xi}). Third, we obtain the first- and
second-order terms in the expansion of the effective action in the radial
displacement of the arc, and reduce the second-order form factor to a
one-loop expression involving the radial Green's functions of each angular
sector, whose imaginary part gives the probability of pair creation in
terms of the radial modes (Sect.~\ref{sec:dyn}). Fourth, we study the limit
of a very narrow wedge, in which each angular sector becomes a massive field
with a moving $\delta$-mirror, the mass being set by the angular
quantization, and the lowest sector dominates the pair creation in a
frequency window (Sect.~\ref{sec:narrow}). We also extend the static results
to the electromagnetic field, for perfect electric or magnetic conductor
walls, and obtain the energy of a perfectly reflecting arc with Neumann
conditions (Sects.~\ref{sec:arcN} and~\ref{sec:em}).

The paper is organized as follows. In Sect.~\ref{sec:themodel} we define
the model; Sect.~\ref{sec:eff} reviews the worldline representation of the
effective action and its perturbative expansion in the radial displacement
of the arc; Sect.~\ref{sec:zero} contains the zeroth-order (static) results,
including the arc at arbitrary $\lambda$, the $\xi=\frac{1}{4}$
identification, and the Neumann arc; Sect.~\ref{sec:dyn} deals with the
first- and second-order terms; Sect.~\ref{sec:narrow} with the narrow-wedge
limit; Sect.~\ref{sec:em} contains the electromagnetic extensions; and
Sect.~\ref{sec:conclusions} presents our conclusions.

\section{The model}\label{sec:themodel}
We consider a real scalar field $\varphi(x)$ in $D = d+1$ dimensions,
coupled to static and dynamical media, in such a way that its Euclidean action 
${\mathcal S}(\varphi, V)$ may be written, after an integration by parts,
as follows:
\begin{equation}\label{eq:action}
{\mathcal S}(\varphi, V) \;=\; \frac{1}{2} \, \int \dd^Dx \, \varphi(x) 
\big(-\partial^2 +  V \big)  \, \varphi(x)  \;.
\end{equation}
We use the notation $x \equiv (x_0, x_1,\dots,x_{d-2}, x_\perp)$, where $x_0$ 
is the imaginary time, $x_1, \dots, x_{d-2}$ are the coordinates along the edge 
of the wedge (the axial coordinates), and $x_\perp \equiv (x_{d-1}, x_d) = 
(\rho\cos\theta, \rho\sin\theta)$ are the coordinates in the plane orthogonal 
to the edge, written in polar form. Natural units are used ($\hbar\equiv 1$, 
$c\equiv 1$), and the metric reduces to the $(d+1)\times(d+1)$ identity matrix, 
so no distinction is implied by placing an index upstairs or downstairs.

The potential is the sum of two terms,
\begin{equation}\label{eq:defV}
V(x) \;=\; V_w(\theta) + V_\lambda(x)\;.
\end{equation}
The first one confines the field to the angular sector $0<\theta<\alpha$,
\begin{equation}\label{eq:defvwedge}
V_w(\theta) \,=\, \left\{\begin{array}{ll} 0\;, & 0<\theta<\alpha \\
+\infty\;, & \text{otherwise}\end{array}\right. \;,
\end{equation}
while the second one describes a semitransparent circular arc, whose radius 
may oscillate around a mean value $a$:
\begin{equation}\label{eq:defVlambda}
V_\lambda(x) \;=\; v\big(\rho-a-R(x_0)\big)\;,\qquad v(u) \,=\, \lambda\,
\delta(u)\;,\qquad \lambda\geq0\;,
\end{equation}
where $R(x_0)$, the Euclidean version of the radial displacement $R(t)$ of the
arc, is treated as a small perturbation. Our aim is to calculate the effective
action in powers of $R$, for an arbitrary value of $\lambda$. Throughout, it
will be convenient to use
\begin{equation}
n \,\equiv\, \frac{\pi}{\alpha}\;,
\end{equation}
which is not restricted to integer values: $n$ is an arbitrary real number,
with $n\geq\frac{1}{2}$ (that is, $\alpha\leq 2\pi$); $n=1$ corresponds to a
single plane and $n=\frac{1}{2}$ to a half-plane.

The boundary conditions on the walls $\theta=0$ and $\theta=\alpha$ define the
domain of the self-adjoint operator $-\partial^2+V$ appearing
in~\eqref{eq:action}. Since the worldline representation follows from writing
the effective action in terms of $\mathrm{Tr}\,e^{-T(-\partial^2+V)}$, the same
operator, with the same domain, plays the role of the Hamiltonian of the
point-like particle; therefore, its propagator obeys the same boundary
conditions as the field. Dirichlet conditions may be obtained as the
strong-coupling limit of a potential~\cite{Gies_WL}, as in~\eqref{eq:defvwedge},
while Neumann (or mixed) conditions, which correspond to paths reflected at the
walls, are implemented directly through the corresponding propagators.
Accordingly, we regard~\eqref{eq:defvwedge} as a shorthand for the restriction
of the field to the sector $0<\theta<\alpha$, the boundary conditions on each
wall (Dirichlet or Neumann) being specified separately.

\section{Effective action}\label{sec:eff}
The one-loop effective action $\Gamma(V)$ is defined in terms of the 
vacuum-to-vacuum Euclidean transition amplitude, ${\mathcal Z}$, via
\begin{equation}
e^{-\Gamma(V)}\,\equiv\, \frac{{\mathcal Z}(V)}{{\mathcal Z}(0)}\;,
\;\;\; {\mathcal Z}(V) \,\equiv \,\int {\mathcal D}\varphi \, 
e^{-{\mathcal S}(\varphi, V)} \;,
\end{equation}
where the ${\mathcal Z}(0)$ factor subtracts the zero-point contribution of 
free space. In the worldline approach~\cite{Gies_WL,Worldline_Review},
\begin{equation}\label{eq:defwl}
 \Gamma(V) = -\frac{1}{2 (4 \pi)^{\frac{d+1}{2}}} \int_{0^+}^\infty 
\frac{\dd T}{T^{\frac{d+3}{2}}} \int \dd^{d+1}x \left[ \langle e^{-\int_0^T \dd\tau 
 V(x(\tau)) } \rangle_x  -1  \right] \;,
\end{equation}
where $\langle \dots \rangle_x$ denotes a functional average over closed loops
$x_\mu(\tau)$ with $x_\mu(0)=x_\mu(T) \equiv x$:
\begin{equation}
\langle \ldots \rangle_x \;\equiv\; 
\frac{\int^{x(T)=x}_{x(0)=x} {\mathcal D}x \ldots e^{-\frac{1}{4}\int_0^T \dd\tau 
\dot{x}_\mu(\tau)\dot{x}_\mu(\tau)}}{\int^{x(T)=x}_{x(0)=x} {\mathcal D}x 
e^{-\frac{1}{4}\int_0^T \dd\tau \dot{x}_\mu(\tau)\dot{x}_\mu(\tau)}} \;.
\end{equation}

Since $V$ depends on $x_0$ and $x_\perp$, but not on the axial coordinates, 
the latter cancel between numerator and denominator, and the effective action 
is proportional to their $(d-2)$-dimensional volume $L^{d-2}$. We therefore 
define $\widehat{\Gamma}\equiv\Gamma/L^{d-2}$:
\begin{equation}\label{eq:gammahat}
\begin{split}
\widehat{\Gamma}(V) \;=\; -\frac{1}{2(4\pi)^{\frac{d+1}{2}}}&\int_{0^+}^\infty
\frac{\dd T}{T^{\frac{d+3}{2}}}\int \dd x_0\int \dd^2x_\perp\\
&\times\Big[
\big\langle e^{-\int_0^T\dd\tau\,V(x_0(\tau),\,x_\perp(\tau))}
\big\rangle_{(x_0,\,x_\perp)}-1\Big]\;,
\end{split}
\end{equation}
where $\int\dd^2x_\perp=\int\dd\theta\int_0^\infty\dd\rho\,\rho$ and 
$\langle\dots\rangle_{(x_0,x_\perp)}$ is the average over closed loops in 
the three coordinates $(x_0,x_\perp)$. We denote by $\langle\dots\rangle_{x_0}$ 
and $\langle\dots\rangle_{x_\perp}$ the averages over closed loops in $x_0$ 
and in $x_\perp$, respectively, each normalized by the corresponding free 
path integral, and write $\rho(\tau)\equiv|x_\perp(\tau)|$.

\subsection{Perturbative expansion}\label{sec:perturbative}
Expanding the potential in powers of $R$,
\begin{equation}\label{eq:Vexp}
V(x) \;=\; V_0(x_\perp) + \sum_{k=1}^\infty\frac{(-1)^k}{k!}\,
v^{(k)}(\rho-a)\,\big(R(x_0)\big)^k\;,
\end{equation}
with $V_0(x_\perp)\equiv V_w(\theta)+v(\rho-a)$,
we have, up to second order,
\begin{align}
e^{-\int_0^T\dd\tau\,V} \;=\; e^{-\int_0^T\dd\tau\,V_0}\,\Big[\,1
&+\int_0^T\dd\tau\;v'\big(\rho(\tau)-a\big)\,R\big(x_0(\tau)\big)
\nonumber\\
&-\frac{1}{2}\int_0^T\dd\tau\;v''\big(\rho(\tau)-a\big)\,
R^2\big(x_0(\tau)\big)
\nonumber\\
&+\int_0^T\dd\tau_1\int_0^{\tau_1}\dd\tau_2\;
v'\big(\rho(\tau_1)-a\big)\,v'\big(\rho(\tau_2)-a\big)\nonumber\\
&\qquad\times R\big(x_0(\tau_1)\big)\,R\big(x_0(\tau_2)\big)
\;+\;O(R^3)\Big]\;.
\end{align}
Since $V_0$ and the derivatives $v^{(k)}$ depend only on $x_\perp$, while the
insertions of $R$ depend only on $x_0$, every term factorizes into an $x_0$
average and an average over loops in the transverse plane. Thus,
$\widehat{\Gamma}=\widehat{\Gamma}_0+\widehat{\Gamma}_1+\widehat{\Gamma}_{2,1}
+\widehat{\Gamma}_{2,2}+O(R^3)$, with:
\begin{equation}\label{eq:G0dyn}
\begin{split}
\widehat{\Gamma}_0 \;=\; -\frac{1}{2(4\pi)^{\frac{d+1}{2}}}&\int_{0^+}^\infty
\frac{\dd T}{T^{\frac{d+3}{2}}}\int\dd x_0\int\dd^2x_\perp\\
&\times\Big[
\big\langle e^{-\int_0^T\dd\tau\,V_0(x_\perp(\tau))}\big\rangle_{x_\perp}-1\Big]
\;,
\end{split}
\end{equation}
\begin{align}\label{eq:G1dyn}
\widehat{\Gamma}_1 \;=\; -\frac{1}{2(4\pi)^{\frac{d+1}{2}}}&\int_{0^+}^\infty
\frac{\dd T}{T^{\frac{d+3}{2}}}\int_0^T\dd\tau\;\nonumber\\
\times&\Big[\int\dd x_0\,\big\langle R\big(x_0(\tau)\big)\big\rangle_{x_0}\Big]
\nonumber\\
\times&\Big[\int\dd^2x_\perp\,\big\langle v'\big(\rho(\tau)-a\big)\,
e^{-\int_0^T\dd\tau'\,V_0(x_\perp(\tau'))}\big\rangle_{x_\perp}\Big]\,,
\end{align}
\begin{align}\label{eq:G21dyn}
\widehat{\Gamma}_{2,1} \;=\; \frac{1}{4(4\pi)^{\frac{d+1}{2}}}&\int_{0^+}^\infty
\frac{\dd T}{T^{\frac{d+3}{2}}}\int_0^T\dd\tau\;\nonumber\\
\times&\Big[\int\dd x_0\,\big\langle R^2\big(x_0(\tau)\big)\big\rangle_{x_0}\Big]
\nonumber\\
\times&\Big[\int\dd^2x_\perp\,\big\langle v''\big(\rho(\tau)-a\big)\,
e^{-\int_0^T\dd\tau'\,V_0(x_\perp(\tau'))}\big\rangle_{x_\perp}\Big]\;,
\end{align}
\begin{align}\label{eq:G22dyn}
\widehat{\Gamma}_{2,2} \;=\; -\frac{1}{2(4\pi)^{\frac{d+1}{2}}}&\int_{0^+}^\infty
\frac{\dd T}{T^{\frac{d+3}{2}}}\int_0^T\dd\tau_1\int_0^{\tau_1}\dd\tau_2\;\nonumber\\
\times&\Big[\int\dd x_0\,\big\langle R\big(x_0(\tau_1)\big)\,R\big(x_0(\tau_2)\big)
\big\rangle_{x_0}\Big]
\nonumber\\
\times&\Big[\int\dd^2x_\perp\,\big\langle v'\big(\rho(\tau_1)-a\big)\,
v'\big(\rho(\tau_2)-a\big)\nonumber\\
&\qquad\times e^{-\int_0^T\dd\tau\,V_0(x_\perp(\tau))}\big\rangle_{x_\perp}\Big]\;.
\end{align}
$\widehat{\Gamma}_{2,1}$ contains a single insertion of the second-order term 
of~\eqref{eq:Vexp}, while $\widehat{\Gamma}_{2,2}$ contains two insertions of 
the first-order one. In all cases, the transverse factor is a closed-loop average
in the static potential $V_0$, i.e., in the presence of the wedge and of the 
arc at its mean position.

\section{Zeroth order: Casimir energy}\label{sec:zero}

\subsection{Vacuum energy}\label{sec:angE}
The zeroth-order term~\eqref{eq:G0dyn} is proportional to the total
evolution time $\mathcal{T}$. We then define ${\mathcal E}$, the vacuum energy
per unit $(d-2)$-dimensional volume of the edge (per unit length of the edge
for $d=3$):
\begin{equation}\label{eq:cale}
\begin{split}
{\mathcal E} \,\equiv\,& \Big[\frac{\widehat{\Gamma}_0}{\mathcal{T}}
\Big]_{\mathcal{T} \to \infty} \\
\;=\;& -\frac{1}{2 (4 \pi)^{\frac{d+1}{2}}} \int_{0^+}^\infty 
\frac{\dd T}{T^{\frac{d+3}{2}}} \int_0^\alpha \dd\theta \int_0^\infty \dd\rho\,\rho 
\left[ (4\pi T)\,K_\perp(T;\,\rho,\theta) -1  \right] \;,
\end{split}
\end{equation}
where we used $\langle e^{-\int_0^T\dd\tau\,V_0}\rangle_{x_\perp} = (4\pi T)\,
K_\perp(T;x_\perp,x_\perp)$, in terms of the closed-loop kernel of a 
two-dimensional particle of mass $1/2$ in the potential $V_0$,
\begin{equation}\label{eq:defKperp}
K_\perp(T;\,x_\perp,y_\perp) \;\equiv\; 
\int_{x_\perp(0)=y_\perp}^{x_\perp(T)=x_\perp} {\mathcal D}x_\perp\;
e^{-\int_0^T \dd\tau\,\big[\frac{1}{4}\dot{x}_\perp^2(\tau) + V_0(x_\perp(\tau))\big]}\;,
\end{equation}
normalized in such a way that, for $V_0=0$, $K_\perp(T;x_\perp,x_\perp) = 
(4\pi T)^{-1}$; $K_\perp(T;\rho,\theta)$ denotes this kernel at coincident 
points.

Equation~\eqref{eq:cale} displays the main practical advantage of the 
worldline approach: the only input it requires is $K_\perp$, the Euclidean 
propagator of a non-relativistic particle with Hamiltonian $p_\perp^2+V_0$ 
(mass $1/2$, $\hbar=1$), or, equivalently, its Laplace transform in $T$, the 
corresponding energy Green's function. Moreover, since $K_\perp$ satisfies 
$\partial_TK_\perp=(\nabla_\perp^2-V_0)K_\perp$, when $V_0$ describes 
impenetrable walls $K_\perp$ is also the Green's function of the heat equation 
with unit diffusivity in the region they enclose. Therefore, whenever these 
objects are available in the literature on non-relativistic path 
integrals~\cite{GroscheSteiner} or on heat conduction~\cite{CarslawJaeger,Cole}, 
we take them from there, and only the proper-time and spatial integrals 
in~\eqref{eq:cale} remain to be performed. Outside the wedge the kernel
vanishes, and only the subtraction contributes; this is why the angular
integral in~\eqref{eq:cale} has been restricted to $0<\theta<\alpha$. The terms
coming from the subtraction are proportional to the volume of the region and
contain no scale: in dimensional regularization, which we use throughout, they
vanish. Equivalently, they correspond to the free-space energy of the region,
which is removed by renormalization.

\subsection{Wedge without arc: Dirichlet walls}\label{sec:dirichlet}
We first consider $\lambda=0$, so that $V_0=V_w$. The kernel of a particle 
confined to a wedge with Dirichlet walls is a known result: it is the 
Green's function of the heat equation for the wedge $0<\theta<\alpha$ with 
vanishing temperature on its faces~\cite{CarslawJaeger}.
Written as a sum over angular modes, at coincident points it reads
\begin{equation}\label{eq:Kwedge}
K_\perp(T;\,\rho,\theta) \;=\; \frac{2}{\alpha}\sum_{\nu=1}^\infty 
\sin^2(n\nu\theta)\int_0^\infty \dd k\,k\;e^{-k^2T}\,\big[J_{n\nu}(k\rho)\big]^2 \;.
\end{equation}
Introducing~\eqref{eq:Kwedge} into~\eqref{eq:cale}, the angular integral gives 
$\int_0^\alpha \dd\theta\,\sin^2(n\nu\theta) = \alpha/2$ for every $\nu\geq1$. 
The proper-time integral is
\begin{equation}
\int_0^\infty \dd T\; T^{-\frac{d+1}{2}}\,e^{-k^2T} \;=\; 
k^{d-1}\,\Gamma\Big(\frac{1-d}{2}\Big)\;,
\end{equation}
and the remaining $k$ integral is of the form
\begin{equation}\label{eq:besselint}
\int_0^\infty \dd k\;k^{s-1}\,\big[J_\ell(k\rho)\big]^2 \;=\; 
\frac{\rho^{-s}}{2\sqrt{\pi}}\;
\frac{\Gamma\big(\frac{1-s}{2}\big)\,\Gamma\big(\ell+\frac{s}{2}\big)}
{\Gamma\big(1-\frac{s}{2}\big)\,\Gamma\big(\ell+1-\frac{s}{2}\big)}\;,
\end{equation}
valid for $\mathrm{Re}\,s<1$ and $\mathrm{Re}(2\ell+s)>0$, which we use with 
$s=d+1$ by analytic continuation. The factor $\Gamma(\frac{1-d}{2})$ produced 
by the proper-time integral cancels against $\Gamma(1-\frac{s}{2})$ in the 
denominator of~\eqref{eq:besselint}. Since both factors become $\Gamma(-1)$ 
at $d=3$, this cancellation is meaningful only for non-integer $d$; 
consistently, every subsequent step must be understood as analytically 
continued in $d$. We obtain
\begin{equation}\label{eq:Ewedge_d}
{\mathcal E}_\mathrm{D} \;=\; -\frac{\sqrt{\pi}\;\Gamma\big(-\frac{d}{2}\big)}
{(4\pi)^{\frac{d+1}{2}}}\; S_n(d)\int_0^\infty \frac{\dd\rho}{\rho^{d}}\;,
\end{equation}
with
\begin{equation}\label{eq:Sn}
S_n(d) \;\equiv\; \sum_{\nu=1}^\infty f_d(n\nu)\;,\qquad 
f_d(x) \;\equiv\; \frac{\Gamma\big(x+\frac{d+1}{2}\big)}{\Gamma\big(x+\frac{1-d}{2}\big)}\;.
\end{equation}

At $d=3$ the summand is a polynomial, $f_3(x) = x(x^2-1)$, and the naive use of 
$\sum_{\nu=1}^\infty \nu^k = \zeta(-k)$ would give $S_n(3)=n^3\zeta(-3)-n\,
\zeta(-1)$. This, however, mixes two different regularization schemes: the 
cancellation of the $\Gamma(\frac{1-d}{2})$ factors above already relies on 
the continuation in $d$, so the sum $S_n(d)$ must be continued in $d$ as well, 
taking the limit $d\to3$ at the end. To that end, we use the large-$x$ 
asymptotic expansion
\begin{equation}
f_d(x) \;\sim\; x^d + c_2(d)\,x^{d-2} + c_4(d)\,x^{d-4} + O(x^{d-6})\;,
\end{equation}
with
\begin{equation}
c_2(d) \,=\, -\frac{d(d^2-1)}{24}\;,\qquad 
c_4(d) \,=\, \frac{d(d+1)(d-1)(d-2)(d-3)(5d+7)}{5760}\;,
\end{equation}
the coefficients of odd powers being zero. Then
\begin{equation}
S_n(d) \;=\; n^d\zeta(-d) + c_2(d)\,n^{d-2}\zeta(2-d) + c_4(d)\,n^{d-4}\zeta(4-d) 
+ R_n(d)\;,
\end{equation}
where
\begin{equation}
R_n(d) \;=\; \sum_{\nu=1}^\infty\Big[f_d(n\nu)-(n\nu)^d-c_2(d)(n\nu)^{d-2}
-c_4(d)(n\nu)^{d-4}\Big]
\end{equation}
is analytic for $\mathrm{Re}\,d<5$ and vanishes at $d=3$, since 
$f_3(x)=x^3-x$, $c_2(3)=-1$ and $c_4(3)=0$. The first two terms reproduce the 
naive result, while the third one gives a finite contribution, because 
$c_4(d) = \frac{11}{120}(d-3) + O\big((d-3)^2\big)$ and 
$\zeta(4-d) = \frac{1}{3-d} + O(1)$:
\begin{equation}
S_n(3) \;=\; \frac{n^3}{120} + \frac{n}{12} - \frac{11}{120\,n} 
\;=\; \frac{(n^2-1)(n^2+11)}{120\,n}\;.
\end{equation}
Since $-\sqrt{\pi}\,\Gamma(-\frac{3}{2})/(4\pi)^2 = -1/(12\pi)$, we arrive at
\begin{equation}\label{eq:ED}
{\mathcal E}_\mathrm{D} \;=\; \int_0^\infty \dd\rho\;\varepsilon_\mathrm{D}(\rho)\;,
\qquad
\varepsilon_\mathrm{D}(\rho) \;=\; -\frac{n^4+10n^2-11}{1440\,\pi\,n}\;
\frac{1}{\rho^3}\;.
\end{equation}
Since the problem has no length scale, the $\rho$ integral is scale-less (it 
vanishes in dimensional regularization, and it diverges at the edge with any 
other regulator). The physical content is in the radial density 
$\varepsilon_\mathrm{D}(\rho)$, the energy per unit length of the edge and per unit 
radial coordinate. Note that it vanishes for $n=1$, when the wedge reduces to 
a single plane.

\subsection{Wedge without arc: Neumann and mixed walls}\label{sec:neumann}
For Neumann conditions on both walls, the kernel is also 
known~\cite{CarslawJaeger}:
it follows from~\eqref{eq:Kwedge} by replacing $\sin\to\cos$, the sum 
including the mode $\nu=0$ with weight $1/\alpha$. After 
the angular integration all modes enter with unit weight, and the only 
difference with respect to the Dirichlet case is the term 
$f_d(0) = \Gamma(\frac{d+1}{2})/\Gamma(\frac{1-d}{2})$, which vanishes at $d=3$. 
Therefore ${\mathcal E}_\mathrm{N} = {\mathcal E}_\mathrm{D}$.

For mixed conditions, Dirichlet at $\theta=0$ and Neumann at $\theta=\alpha$, 
the angular modes are $\sin\big(n(\nu+\frac{1}{2})\theta\big)$,
$\nu=0,1,2,\dots$, namely, the odd modes of a Dirichlet wedge of opening angle
$2\alpha$. Hence the kernel follows from the Dirichlet one through the identity
$K^{\alpha}_\mathrm{DN}=2K^{2\alpha}_\mathrm{D}-K^{\alpha}_\mathrm{D}$, valid for
arbitrary real $n$, which amounts to replacing $n\nu\to n(\nu+\frac{1}{2})$
in~\eqref{eq:Kwedge}; the angular integral is again $\alpha/2$.
Hence~\eqref{eq:Ewedge_d} holds with $S_n(d)$ replaced by 
$\widetilde{S}_n(d) = \sum_{\nu=0}^\infty f_d\big(n(\nu+\frac{1}{2})\big)$. The 
continuation follows the same steps, with $\zeta(s)\to\zeta(s,\frac{1}{2}) = 
(2^s-1)\zeta(s)$; the $c_4$ term again contributes $-\frac{11}{120n}$, and
\begin{equation}
\widetilde{S}_n(3) \;=\; -\frac{7n^4+40n^2+88}{960\,n} \;=\; S_{n/2}(3)-S_n(3)\;,
\end{equation}
where the last equality is the same identity, now at the level of the mode
sums. Thus
\begin{equation}\label{eq:EDN}
\varepsilon_\mathrm{DN}(\rho) \;=\; \frac{7n^4+40n^2+88}{11520\,\pi\,n}\;
\frac{1}{\rho^3}\;,
\end{equation}
which is positive for all $n$.

\subsection{Semitransparent arc}\label{sec:arcgreen}
We now consider $\lambda>0$, with the arc at its mean position, so that 
$V_0(x_\perp)=V_w(\theta)+\lambda\,\delta(\rho-a)$. Since the new term does not depend on $\theta$, the angular decomposition 
of~\eqref{eq:Kwedge} is unchanged, and for Dirichlet walls
\begin{equation}
K_\perp(T;\,\rho,\theta) \;=\; \frac{2}{\alpha}\sum_{\nu=1}^\infty 
\sin^2(\mu_\nu\theta)\;k_\nu(T;\,\rho,\rho)\;,\qquad \mu_\nu\equiv n\nu\;,
\end{equation}
where $k_\nu$ is the kernel of the radial operator
\begin{equation}
h_\nu \;=\; -\partial_\rho^2-\frac{1}{\rho}\,\partial_\rho+\frac{\mu_\nu^2}{\rho^2}
+\lambda\,\delta(\rho-a)
\end{equation}
acting on $L^2(\rho\,\dd\rho)$. For Neumann or mixed walls, $\mu_\nu$ is 
replaced by the corresponding angular orders. The proper-time transform of
$k_\nu$,
\begin{equation}
{\mathcal G}_\nu(E;\,\rho,\rho') \;\equiv\; \int_0^\infty\dd t\;e^{-E\,t}\,
k_\nu(t;\,\rho,\rho') \;=\; \bra{\rho}\big(h_\nu+E\big)^{-1}\ket{\rho'}\;,
\end{equation}
is, for $\lambda=0$ and with $E=\kappa^2$, the well-known radial Green's 
function of the free particle~\cite{GroscheSteiner},
\begin{equation}
{\mathcal G}^{(0)}_\nu(E;\,\rho,\rho') \;=\; I_{\mu_\nu}(\kappa\rho_<)\,
K_{\mu_\nu}(\kappa\rho_>)\;,
\end{equation}
where $\rho_<=\min(\rho,\rho')$, $\rho_>=\max(\rho,\rho')$, and $I_\mu$, 
$K_\mu$ are modified Bessel functions. Path integrals with $\delta$-function 
perturbations are treated in Ref.~\cite{GroscheSteiner}, where the Green's 
function in the presence of the $\delta$ term is given in terms of the one 
without it; here, the integration measure $\rho\,\dd\rho$ produces a factor 
$a$, and
\begin{equation}\label{eq:Gdeltaarc}
{\mathcal G}_\nu(E;\,\rho,\rho') \;=\; {\mathcal G}^{(0)}_\nu(E;\,\rho,\rho')
-\frac{\lambda a\;{\mathcal G}^{(0)}_\nu(E;\,\rho,a)\,{\mathcal G}^{(0)}_\nu(E;\,a,\rho')}
{1+\lambda a\;I_{\mu_\nu}(\kappa a)\,K_{\mu_\nu}(\kappa a)}\;.
\end{equation}
For $\lambda\geq0$ the denominator has no zeros, so there are no bound states, 
and the kernel follows from the inverse transform,
\begin{equation}
k_\nu(t;\,\rho,\rho') \;=\; \frac{1}{2\pi\ii}\int_{c-\ii\infty}^{c+\ii\infty}\dd E\;
e^{E\,t}\;{\mathcal G}_\nu(E;\,\rho,\rho')\;,\qquad c>0\;.
\end{equation}

Splitting ${\mathcal E} = {\mathcal E}_0 + {\mathcal E}_\lambda$, where 
${\mathcal E}_0$ is the energy of the wedge without the arc, the new term is
\begin{align}\label{eq:Elambda_T}
{\mathcal E}_\lambda \;=\;& \frac{1}{2(4\pi)^{\frac{d-1}{2}}}\int_{0^+}^\infty
\frac{\dd T}{T^{\frac{d+1}{2}}}\int_0^\alpha\dd\theta\int_0^\infty\dd\rho\,\rho\;
\frac{1}{2\pi\ii}\int_{c-\ii\infty}^{c+\ii\infty}\dd E\;e^{E\,T}
\nonumber\\
&\times\frac{2}{\alpha}\sum_{\nu=1}^\infty\sin^2(\mu_\nu\theta)\;
\frac{\lambda a\,\big[I_{\mu_\nu}(\kappa\rho_<)\,K_{\mu_\nu}(\kappa\rho_>)\big]^2}
{1+\lambda a\,I_{\mu_\nu}(\kappa a)\,K_{\mu_\nu}(\kappa a)}\;,
\end{align}
where now $\rho_<=\min(\rho,a)$ and $\rho_>=\max(\rho,a)$. The proper-time 
integral is a Mellin transform,
\begin{equation}
\int_0^\infty\dd T\;T^{-s}\,k_\nu(T) \;=\; \frac{1}{\Gamma(s)}\int_0^\infty\dd E\;
E^{s-1}\,{\mathcal G}_\nu(E)\;,\qquad s=\frac{d+1}{2}\;,
\end{equation}
and, since the perturbation is of rank one in the radial variable, the 
$\rho$ integral may be performed through the trace identity
\begin{equation}
\int_0^\infty\dd\rho\,\rho\;\Big[{\mathcal G}_\nu(E;\,\rho,\rho)
-{\mathcal G}^{(0)}_\nu(E;\,\rho,\rho)\Big] \;=\; 
\partial_E\log\Big[1+\lambda a\,I_{\mu_\nu}(\kappa a)\,K_{\mu_\nu}(\kappa a)\Big]\;.
\end{equation}
Therefore,
\begin{equation}\label{eq:Elambda}
\begin{split}
{\mathcal E}_\lambda \;=\;& -\frac{2\pi}{(4\pi)^{\frac{d+1}{2}}\,
\Gamma\big(\frac{d+1}{2}\big)}\\
&\times\sum_{\nu=1}^\infty\int_0^\infty\dd E\;
E^{\frac{d-1}{2}}\;\partial_E\log\Big[1+\lambda a\,I_{\mu_\nu}(\sqrt{E}a)\,
K_{\mu_\nu}(\sqrt{E}a)\Big]\;,
\end{split}
\end{equation}
to be evaluated, as before, by analytic continuation in $d$. This expression 
is exact in $\lambda$. Its expansion in powers of $\lambda$ connects with 
perturbative calculations, while the expansion in powers of $1/\lambda$ follows 
from $\log(1+u) = \log u + \sum_{m\geq1}\frac{(-1)^{m+1}}{m\,u^m}$, with 
$u = \lambda a\,I_{\mu_\nu}K_{\mu_\nu}$. Since $I_\mu(\kappa a)K_\mu(\kappa a) 
\simeq (2\kappa a)^{-1}$ for large $\kappa$, the latter is not uniform in 
$\kappa$, and must be understood as an asymptotic expansion. Note also that
${\mathcal E}_0$ does not depend on $a$: the whole dependence on the radius of
the arc, and hence the radial force acting on it, is contained in
${\mathcal E}_\lambda$.

In the Dirichlet limit, $\lambda\to\infty$, the term $\log\lambda a$ does not depend on $E$ 
and drops out, so that
\begin{equation}\label{eq:EDarc}
{\mathcal E}^\mathrm{arc}_\mathrm{D} \;=\; -\frac{2\pi\,a^{1-d}}{(4\pi)^{\frac{d+1}{2}}\,
\Gamma\big(\frac{d+1}{2}\big)}\sum_{\nu=1}^\infty\int_0^\infty\dd w\;w^{d-1}\,
\partial_w\log\Big[I_{n\nu}(w)\,K_{n\nu}(w)\Big]\;,
\end{equation}
where $\log I_{n\nu}$ corresponds to the interior region, $\rho<a$ (the wedge 
closed by a Dirichlet arc), and $\log K_{n\nu}$ to the exterior one. Indeed, 
the Green's function~\eqref{eq:Gdeltaarc} tends to
\begin{equation}
{\mathcal G}_\nu \;\to\; I_{\mu_\nu}(\kappa\rho_<)K_{\mu_\nu}(\kappa\rho_>)-
\left\{\begin{array}{ll}
I_{\mu_\nu}(\kappa\rho)\,I_{\mu_\nu}(\kappa\rho')\,
\dfrac{K_{\mu_\nu}(\kappa a)}{I_{\mu_\nu}(\kappa a)}\;, & \rho,\rho'<a\;,\\[10pt]
K_{\mu_\nu}(\kappa\rho)\,K_{\mu_\nu}(\kappa\rho')\,
\dfrac{I_{\mu_\nu}(\kappa a)}{K_{\mu_\nu}(\kappa a)}\;, & \rho,\rho'>a\;,
\end{array}\right.
\end{equation}
which vanishes at $\rho=a$. These are the Green's functions of the interior 
and exterior Dirichlet problems for the circular boundary; for the interior 
sector, the corresponding series representation of the kernel can be found in 
Ref.~\cite{CarslawJaeger}.

We note that, unlike the case of the wedge alone, the arc introduces a
curved boundary and, at each of its ends, two right-angled corners (one on
each side of the arc) where it meets the wall. A uniform asymptotic expansion
of the logarithm in~\eqref{eq:EDarc} shows that the only possible pole at
$d=3$ comes from the term that behaves as $w^{-2}$ at large $w$, and that its
residue is proportional to the zeta-regularized number of angular orders,
$\sum_\nu\mu_\nu^{0}$, an order $\mu_\nu=0$ counting as one. For Dirichlet
walls this number is $\zeta(0)=-\frac{1}{2}$, for Neumann walls it is
$1+\zeta(0)=+\frac{1}{2}$, and in both cases a logarithmic divergence
associated with the corners appears. It does not depend on $\alpha$, so that
the $\alpha$-dependence of ${\mathcal E}^\mathrm{arc}_\mathrm{D}$, and therefore the
torque on the walls, is finite, while the radial force on the arc is not. For
mixed walls the number of orders is $\zeta(0,\frac{1}{2})=0$, and the energy is
finite. For the full circular cylinder (Dirichlet and Neumann walls at $n=1$
combined) the two residues cancel, and the same procedure reproduces the known
value $0.000614794/a^2$ (Eq.~(34) of Ref.~\cite{GosdzinskyRomeo}).

\subsection{Comparison with the literature: the \texorpdfstring{$\xi=1/4$}{xi = 1/4} density}\label{sec:xi}

The energies obtained so far are global quantities, defined from the 
effective action. In the literature, the vacuum energy inside the wedge has 
been obtained by local methods, namely, as the vacuum expectation value of the 
energy-momentum tensor (EMT). For a scalar field, the energy density depends 
on the curvature coupling $\xi$ of the improved EMT, $T^{00}_\xi = T^{00}_0 
- \xi\,\nabla^2\varphi^2$, with $T^{00}_0 = \frac{1}{2}\dot{\varphi}^2
+\frac{1}{2}(\nabla\varphi)^2$, and different values of $\xi$ differ by a total 
derivative. A meaningful comparison therefore requires identifying which 
density corresponds to the worldline integrand.

For the wedge without arc, the result usually quoted in the literature is that for the conformally coupled scalar ($\xi=1/6$) with Dirichlet 
conditions~\cite{DC,FMT},
\begin{equation}\label{eq:DCscalar}
\langle T^{00}\rangle_{1/6} \;=\; -\frac{n^4-1}{1440\,\pi^2\rho^4}\;,
\end{equation}
which is independent of $\theta$. Integrating over the angle, the corresponding 
radial density is $-\frac{n^4-1}{1440\pi n}\rho^{-3}$, which does not coincide 
with~\eqref{eq:ED}: subtracting it from $\varepsilon_\mathrm{D}(\rho)$ leaves
$-\frac{n^2-1}{144\pi n}\rho^{-3}$. This difference is
proportional to $n-\frac{1}{n}$, so it cannot be absorbed into the 
subtraction of the free-space energy, which is linear in $\alpha$.

The origin of the discrepancy is the choice of energy density. Let 
$A=-\nabla^2$ be the spatial operator in the cavity, with eigenfunctions 
$u_k$ and eigenvalues $\omega_k^2$. Writing the full spatial kernel as 
$K(T;\vec{x},\vec{x}) = (4\pi T)^{-\frac{d-2}{2}}K_\perp(T;x_\perp,x_\perp)$, the 
integrand of~\eqref{eq:cale} at a point $\vec{x}$ is
\begin{equation}\label{eq:localWL}
e(\vec{x}) \;=\; -\frac{1}{2\sqrt{4\pi}}\int_0^\infty\dd T\;T^{-\frac{3}{2}}\,
K(T;\vec{x},\vec{x}) \;=\; \frac{1}{2}\sum_k \omega_k\,u_k^2(\vec{x})
\;=\; \frac{1}{2}\bra{\vec{x}}A^{\frac{1}{2}}\ket{\vec{x}}\;,
\end{equation}
where we used $\int_0^\infty\dd T\,T^{-3/2}e^{-\omega^2T} = 
\Gamma(-\frac{1}{2})\,\omega$, and the free-space subtraction is understood.
On the other hand, expanding the field in modes in $T^{00}_0$ and using
$|\nabla u_k|^2 = \frac{1}{2}\nabla^2(u_k^2)+\omega_k^2u_k^2$, one finds
$\langle T^{00}_0\rangle = e(\vec{x}) + \frac{1}{4}\nabla^2
\langle\varphi^2\rangle$, so that
\begin{equation}\label{eq:xi14}
e(\vec{x}) \;=\; \langle T^{00}_{1/4}\rangle\;,\qquad 
T^{00}_{1/4} \;=\; \frac{1}{2}\big(\dot{\varphi}^2 - \varphi\,\nabla^2\varphi\big)\;.
\end{equation}
The same identification follows from the local zeta-function representation 
of Ref.~\cite{FP1}, where the regularized energy density contains a 
gradient term proportional to $\frac{1}{4}-\xi$, which vanishes for 
$\xi=1/4$ and leaves precisely the diagonal kernel of $\frac{1}{2}A^{1/2}$. In 
other words, the worldline integrand is, point by point, the energy density
with $\xi=1/4$, the only value for which the density contains no product of
spatial gradients of the field. Its integral reproduces the global energy
$\frac{1}{2}\mathrm{Tr}\,A^{1/2}$ with no boundary term, while for any other value 
of $\xi$ the integrated density differs from it by a term proportional to 
$\frac{1}{4}-\xi$~\cite{FP1}.

This can be checked explicitly. The energy density inside the Dirichlet wedge 
for arbitrary $\xi$ is~\cite{RS}
\begin{align}\label{eq:RSxi}
\langle T^{00}\rangle_\xi \;=\;& -\frac{n^4-1}{1440\pi^2\rho^4}
-\Big(\xi-\frac{1}{6}\Big)\frac{n^2-1}{12\pi^2\rho^4}
\nonumber\\
&+\Big(\xi-\frac{1}{6}\Big)\left[\frac{3n^4}{8\pi^2\rho^4\sin^4(n\theta)}
-\frac{n^2(n^2-1)}{4\pi^2\rho^4\sin^2(n\theta)}\right]\;.
\end{align}
The angular integrals of $\sin^{-2}(n\theta)$ and $\sin^{-4}(n\theta)$ over 
$(\epsilon,\alpha-\epsilon)$ are expressed in terms of $\cot(n\epsilon)$, and 
contain only odd powers of $\epsilon$: they are pure divergences localized on 
the walls, with no finite part, and they are discarded in dimensional 
regularization. Only the $\theta$-independent terms survive, and
\begin{equation}\label{eq:Exi}
\int_0^\alpha\dd\theta\;\rho\,\langle T^{00}\rangle_\xi \;=\; 
-\left[\frac{n^4-1}{1440\pi n}+\Big(\xi-\frac{1}{6}\Big)\frac{n^2-1}{12\pi n}
\right]\frac{1}{\rho^3}\;.
\end{equation}
For $\xi=1/6$ this is the integral of~\eqref{eq:DCscalar}, while for $\xi=1/4$ 
it reproduces exactly~\eqref{eq:ED}. The $\xi$-dependent term is the integral 
of $\nabla^2\langle\varphi^2\rangle$, with
\begin{equation}
\langle\varphi^2\rangle \;=\; \frac{1}{\rho^2}\left[\frac{n^2-1}{48\pi^2}
-\frac{n^2}{16\pi^2\sin^2(n\theta)}\right]\;.
\end{equation}
Its $\theta$-dependent part produces only power divergences on the walls, 
whereas the constant part gives a total derivative in $\rho$, namely, a surface 
term located at the edge $\rho\to0$, where the Dirichlet condition does not 
hold and $\langle\varphi^2\rangle\propto\rho^{-2}$ does not vanish. This is the
reason why, unlike for parallel plates, the radial energy density of the
wedge, and hence the energy contained in any region $r_1<\rho<r_2$, depends on
the choice of~$\xi$.

For mixed conditions, the energy density for conformal coupling, 
$\langle T^{00}\rangle_{1/6} = \frac{7n^4+8}{11520\pi^2\rho^4}$, was obtained 
in Ref.~\cite{FP2}. 
The $\xi=1/4$ density follows from the identity $G^{\alpha}_\mathrm{DN} = 
2\,G^{2\alpha}_\mathrm{D}-G^{\alpha}_\mathrm{D}$ between Green's functions, which 
gives
\begin{align}\label{eq:TDN}
\langle T^{00}\rangle^\mathrm{DN}_{1/4} \;=\;& \frac{7n^4+40n^2+88}{11520\pi^2\rho^4}
+\frac{n^2(n^2-1)}{48\pi^2\rho^4\sin^2(n\theta)}
-\frac{n^4}{32\pi^2\rho^4\sin^4(n\theta)}
\nonumber\\
&-\frac{n^2(n^2-4)}{384\pi^2\rho^4\sin^2(\frac{n\theta}{2})}
+\frac{n^4}{256\pi^2\rho^4\sin^4(\frac{n\theta}{2})}\;.
\end{align}
This expression tends to $\pm\frac{1}{32\pi^2 y^4}$ at a distance $y$ from
the Dirichlet and Neumann walls, respectively, its conformal part agrees with 
Ref.~\cite{FP2}, and for $n=2$ it reproduces their result for the rectangular 
wedge. The $\theta$-dependent terms have no finite part after angular 
integration, and the integral of~\eqref{eq:TDN} coincides 
with~\eqref{eq:EDN}.

We now turn to the Dirichlet arc. The interior problem has been studied in Ref.~\cite{RS}, where the energy 
density is written as the density of the wedge plus a contribution 
$\langle T^{00}\rangle_c$ induced by the arc. The local integrand of our 
expression may be obtained by keeping the $\theta$ and $\rho$ dependence before 
integration. Performing the proper-time integral through the Mellin transform, 
we find, for $\rho<a$,
\begin{equation}\label{eq:ecWL}
e_c(\rho,\theta) \;=\; \frac{2}{(4\pi)^{\frac{d-1}{2}}\,
\Gamma\big(\frac{d+1}{2}\big)\,\alpha\,a^{d+1}}\sum_{\nu=1}^\infty
\sin^2(n\nu\theta)\int_0^\infty\dd z\;z^d\,
\frac{K_{n\nu}(z)}{I_{n\nu}(z)}\,I_{n\nu}^2\Big(z\frac{\rho}{a}\Big)\;.
\end{equation}
In Ref.~\cite{RS}, where $D$ denotes the number of spatial dimensions (our
$d$, not the spacetime dimension $D=d+1$ used above),
$\langle T^{00}\rangle_c$ for arbitrary $\xi$ (Eqs.~(31) and (32) there) contains an overall factor $4\xi-1$, multiplying the functions
$B_l^{(\pm)}$, which are combinations of $I_l'^2$, $I_l^2$ and $l^2I_l^2/y^2$
plus the term $2I_l^2/[(D-1)(4\xi-1)]$. At $\xi=1/4$ only the latter survives,
and, using $(D-1)\Gamma(\frac{D-1}{2}) = 2\Gamma(\frac{D+1}{2})$, their result
becomes identical to~\eqref{eq:ecWL}, for arbitrary dimension $D=d$, with 
$\phi_0=\alpha$, $l=n\nu$ and $x=\rho/a$. Together with the 
agreement~\eqref{eq:Exi} for the wedge part, the complete interior energy 
density coincides with the worldline integrand, point by point, for $\xi=1/4$. 
The surface behavior is consistent with this identification: near the arc and 
away from the walls, the density of Ref.~\cite{RS} behaves as 
$\frac{3(\xi-\xi_c)}{8\pi^2(a-\rho)^4}$ for $D=3$ (with $\xi_c=\frac{D-1}{4D}$
the conformal coupling in their notation, $\xi_c=\frac{1}{6}$ at $D=3$), which for $\xi=1/4$ is 
$\frac{1}{32\pi^2(a-\rho)^4}$, the result for a single Dirichlet plane, 
while the leading divergence at the corners where the arc meets the walls is 
proportional to $4\xi-1$ and vanishes. Since the density diverges near the 
arc, the integrated energy requires the same treatment of surface 
divergences as before.

\subsection{Neumann arc}\label{sec:arcN}
The Dirichlet arc has been obtained as the strong-coupling limit of a
$\delta$-potential. An analogous treatment of a Neumann arc would require a
$\delta'$-type interaction, which is not well defined without a further
prescription, and we leave its study for future work. For a perfect Neumann
arc, the energy may instead be obtained by plugging the corresponding
propagator, which, as in the previous cases, is a known result,
into~\eqref{eq:cale}. For the interior sector, the 
kernel is the Green's function of the heat equation in a cylindrical sector 
with an insulated (Neumann) boundary at $\rho=a$, given in Ref.~\cite{Cole} for 
Neumann walls;
other wall conditions only change the angular orders $\mu_\nu$. In the energy 
representation, the radial Green's function reads, for both regions,
\begin{equation}\label{eq:GNarc}
{\mathcal G}^\mathrm{N}_\nu \;=\; I_{\mu_\nu}(\kappa\rho_<)K_{\mu_\nu}(\kappa\rho_>)-
\left\{\begin{array}{ll}
I_{\mu_\nu}(\kappa\rho)\,I_{\mu_\nu}(\kappa\rho')\,
\dfrac{K'_{\mu_\nu}(\kappa a)}{I'_{\mu_\nu}(\kappa a)}\;, & \rho,\rho'<a\;,\\[10pt]
K_{\mu_\nu}(\kappa\rho)\,K_{\mu_\nu}(\kappa\rho')\,
\dfrac{I'_{\mu_\nu}(\kappa a)}{K'_{\mu_\nu}(\kappa a)}\;, & \rho,\rho'>a\;.
\end{array}\right.
\end{equation}
Proceeding as in Sect.~\ref{sec:arcgreen}, the angular integral and the Mellin 
transform lead to
\begin{equation}
{\mathcal E}^\mathrm{arc}_\mathrm{N} \;=\; -\frac{2\pi}{(4\pi)^{\frac{d+1}{2}}\,
\Gamma\big(\frac{d+1}{2}\big)}\sum_\nu\int_0^\infty\dd E\;E^{\frac{d-1}{2}}\,
\Delta_\nu(E)\;,
\end{equation}
with
\begin{equation}
\Delta_\nu(E) \,\equiv\, \int_0^\infty\dd\rho\,\rho\,
\Big[{\mathcal G}^\mathrm{N}_\nu(E;\,\rho,\rho)-{\mathcal G}^{(0)}_\nu(E;\,\rho,\rho)
\Big]\;.
\end{equation}
With $x=\kappa a$, the radial integrals are given by the Lommel formulas
\begin{align}
\int_0^x \dd t\,t\,I_\mu^2(t) &= \frac{x^2}{2}\Big[\Big(1+\frac{\mu^2}{x^2}\Big)
I_\mu^2(x)-I_\mu'^2(x)\Big]\;,\nonumber\\
\int_x^\infty \dd t\,t\,K_\mu^2(t) &= \frac{x^2}{2}\Big[K_\mu'^2(x)-\Big(1+
\frac{\mu^2}{x^2}\Big)K_\mu^2(x)\Big]\;,
\end{align}
so that, after the cancellation of the terms proportional to 
$I'_\mu K'_\mu$,
\begin{equation}
\Delta_\nu(E) \;=\; \frac{a^2}{2}\Big(1+\frac{\mu_\nu^2}{x^2}\Big)\,
\frac{\big(K_{\mu_\nu}I'_{\mu_\nu}-I_{\mu_\nu}K'_{\mu_\nu}\big)
\big(K_{\mu_\nu}I'_{\mu_\nu}+I_{\mu_\nu}K'_{\mu_\nu}\big)}
{I'_{\mu_\nu}K'_{\mu_\nu}}\;,
\end{equation}
all functions being evaluated at $x$. Using the Wronskian 
$K_\mu I'_\mu-I_\mu K'_\mu=1/x$, the identity
\begin{equation}
\Big(1+\frac{\mu^2}{x^2}\Big)\big(I_\mu K_\mu\big)' \;=\; 
\big(I'_\mu K'_\mu\big)'+\frac{2}{x}\,I'_\mu K'_\mu\;,
\end{equation}
which follows from the Bessel equation, and $\dd x/\dd E = a^2/(2x)$, we 
obtain
\begin{equation}
\Delta_\nu(E) \;=\; \partial_E\log\Big[-E\,I'_{\mu_\nu}(\sqrt{E}a)\,
K'_{\mu_\nu}(\sqrt{E}a)\Big]\;.
\end{equation}
The same steps for the Dirichlet arc give $\partial_E\log[I_{\mu_\nu}
K_{\mu_\nu}]$, in agreement with the result of Sect.~\ref{sec:arcgreen}, 
obtained there from the $\lambda\to\infty$ limit. Therefore,
\begin{equation}\label{eq:ENarc}
{\mathcal E}^\mathrm{arc}_\mathrm{N} \;=\; -\frac{2\pi\,a^{1-d}}{(4\pi)^{\frac{d+1}{2}}\,
\Gamma\big(\frac{d+1}{2}\big)}\sum_\nu\int_0^\infty\dd w\;w^{d-1}\,
\partial_w\log\Big[-w^2\,I'_{\mu_\nu}(w)\,K'_{\mu_\nu}(w)\Big]\;,
\end{equation}
where the sum runs over the angular orders corresponding to the conditions on 
the walls, and the splitting $\log[w\,I'_{\mu_\nu}]+\log[-w\,K'_{\mu_\nu}]$ 
separates the interior and exterior contributions. The factor $w^2$ inside the 
logarithm contributes a scale-less term, which vanishes in dimensional 
regularization. For $\mu_\nu=0$ one has $I'_0=I_1$, so that the interior modes
of that sector (the zeros of $J_1$) are correctly accounted for; the
transverse zero mode, constant over the closed sector, contributes a free
field in the remaining $d-1$ dimensions, which is scale-less and vanishes in
dimensional regularization. In any case, that mode must be excluded for the
TE scalar, since it carries no field.

\section{First and second order: dissipative effects}\label{sec:dyn}

For $R(x_0)\neq0$ the configuration depends on time, and $\widehat{\Gamma}$ is no 
longer proportional to $\mathcal{T}$. The terms of first and second order in 
$R$ describe the response of the vacuum to the motion of the arc. The 
effective action defined above is an in-out quantity: after rotation to 
Minkowski spacetime, its imaginary part determines the vacuum persistence 
amplitude, $|\langle0_\mathrm{out}|0_\mathrm{in}\rangle|^2 = e^{-2\,\mathrm{Im}\,
\widehat{\Gamma}^{(M)}L^{d-2}}$, and, to the lowest (second) order in $R$,
$2\,\mathrm{Im}\,\Gamma^{(M)}=2\,\mathrm{Im}\,\widehat{\Gamma}^{(M)}L^{d-2}$ is the
total probability of creating a pair of quanta. These dissipative effects are
the object of this section. Note that, unlike in the static case, we shall not
compare with vacuum expectation values of the EMT: for time-dependent
configurations, the expectation value of a local operator in the initial vacuum
is an in-in quantity, given by the Schwinger-Keldysh formalism, and has no
direct counterpart in the in-out effective action, whereas the imaginary part
of the latter is a global, well-defined observable.

As in Refs.~\cite{FG_Dirichlet,FG_Neumann,FG_WL}, we write the result in 
terms of the Fourier transform $\widetilde{R}(k_0) \equiv \int\dd x_0\,
e^{-\ii k_0x_0}R(x_0)$, and perform the Wick rotation through the 
continuation $k_0\to-\ii(\omega+\ii0)$, together with $-\widehat{\Gamma}\to
\ii\widehat{\Gamma}^{(M)}$. This prescription fixes the branch of the 
non-analytic terms, which are the only ones that contribute to the imaginary 
part; terms that are polynomial in $k_0^2$, as well as local terms, do not.

\subsection{First order}\label{sec:first}
The $x_0$ average in~\eqref{eq:G1dyn} is
\begin{equation}
\int\dd x_0\,\big\langle R\big(x_0(\tau)\big)\big\rangle_{x_0} \;=\; 
\widetilde{R}(0) \;=\; \int\dd x_0\,R(x_0)\;,
\end{equation}
independently of $\tau$. This 
term is real, and it vanishes when $a$ is chosen as the mean radius of the 
arc, a choice we adopt in what follows. Therefore, $\widehat{\Gamma}_1=0$, and 
there is no dissipative effect at first order.

\subsection{Second order}\label{sec:second}
The $x_0$ averages in~\eqref{eq:G21dyn} and~\eqref{eq:G22dyn} follow from the
worldline correlation function
\begin{equation}
-\frac{1}{2}\big\langle \big[x_0(\tau_1)-x_0(\tau_2)\big]^2 \big\rangle \;=\;
\Delta(\tau_1 - \tau_2) \;,
\qquad \Delta(\tau_1 - \tau_2) \,=\,\frac{(\tau_1 - \tau_2)^2}{T} -
|\tau_1 - \tau_2|  \;,
\end{equation}
which coincides, up to terms that cancel in this combination, with
$\langle x_0(\tau_1)x_0(\tau_2)\rangle$ in the string-inspired
convention~\cite{Worldline_Review}. After integration over the base point
$x_0$, they read
\begin{equation}
\int\dd x_0\,\big\langle R\big(x_0(\tau_1)\big)\,R\big(x_0(\tau_2)\big)
\big\rangle_{x_0} \;=\; \int\frac{\dd k_0}{2\pi}\,
\big|\widetilde{R}(k_0)\big|^2\,e^{\Delta(\tau_1-\tau_2)\,k_0^2}\;,
\end{equation}
which, at coincident proper times, reduces to $\int\dd x_0\,R^2(x_0)$.

Therefore, $\widehat{\Gamma}_{2,1}$ is a local term,
\begin{equation}
\widehat{\Gamma}_{2,1} \;=\; \frac{1}{2}\,m^2\int\dd x_0\,\big(R(x_0)\big)^2
\;,
\end{equation}
with
\begin{equation}
m^2 \;=\; \frac{1}{2(4\pi)^{\frac{d+1}{2}}}\int_{0^+}^\infty
\frac{\dd T}{T^{\frac{d+3}{2}}}\int_0^T\dd\tau\int\dd^2x_\perp\,
\big\langle v''\big(\rho(\tau)-a\big)\,e^{-\int_0^T\dd\tau'\,V_0}
\big\rangle_{x_\perp}\;,
\end{equation}
which renormalizes the restoring force of the arc and cannot generate an 
imaginary part. The dissipative effects are contained in
\begin{equation}\label{eq:G22gamma}
\widehat{\Gamma}_{2,2} \;=\; \frac{1}{2}\int\frac{\dd k_0}{2\pi}\;
\gamma(k_0)\;\big|\widetilde{R}(k_0)\big|^2\;,
\end{equation}
with the form factor
\begin{equation}\label{eq:gammadyn}
\gamma(k_0) \;=\; -\frac{1}{(4\pi)^{\frac{d+1}{2}}}\int_{0^+}^\infty
\frac{\dd T}{T^{\frac{d+3}{2}}}\int_0^T\dd\tau_1\int_0^{\tau_1}\dd\tau_2\;
I_\perp(\tau_1,\tau_2;T)\;e^{\Delta(\tau_1-\tau_2)\,k_0^2}\;,
\end{equation}
where
\begin{equation}\label{eq:Iperp}
I_\perp(\tau_1,\tau_2;T) \;\equiv\; \int\dd^2x_\perp\,\big\langle 
v'\big(\rho(\tau_1)-a\big)\,v'\big(\rho(\tau_2)-a\big)\,
e^{-\int_0^T\dd\tau\,V_0(x_\perp(\tau))}\big\rangle_{x_\perp}\;.
\end{equation}

In terms of the static kernel~\eqref{eq:defKperp}, and with the proper-time 
intervals $T_a=T-\tau_1$, $T_b=\tau_1-\tau_2$, $T_c=\tau_2$,
\begin{equation}
\begin{split}
I_\perp \;=\; (4\pi T)\,\lambda^2\int\dd^2x\,\dd^2y\,\dd^2z\;&
K_\perp(T_a;x,y)\,\delta'(\rho_y-a)\\
&\times K_\perp(T_b;y,z)\,\delta'(\rho_z-a)\,
K_\perp(T_c;z,x)\;.
\end{split}
\end{equation}

The proper-time structure of~\eqref{eq:gammadyn} simplifies considerably.
The integral over the base point $x$ of the transverse loop may be performed with the composition property of the kernel,
$\int\dd^2x\,K_\perp(T_c;z,x)K_\perp(T_a;x,y)=K_\perp(T_a+T_c;z,y)$, so that
$I_\perp$ depends on $\tau_1$ and $\tau_2$ only through $u\equiv\tau_1-\tau_2$.
Moreover, for Dirichlet walls,
$K_\perp(t;y,z)=\frac{2}{\alpha}\sum_\nu\sin(\mu_\nu\theta_y)\sin(\mu_\nu\theta_z)\,
k_\nu(t;\rho_y,\rho_z)$, and the orthogonality of the angular modes reduces
$I_\perp$ to a sum over the angular orders:
\begin{equation}\label{eq:Iperpnu}
I_\perp \;=\; (4\pi T)\,\lambda^2\sum_{\nu}\partial_\rho\partial_{\rho'}
\Big[\rho\rho'\,k_\nu(u;\rho,\rho')\,k_\nu(T-u;\rho',\rho)\Big]_{\rho=\rho'=a}\;.
\end{equation}
The radial kernels have a discontinuous first derivative at $\rho=a$, and
the derivatives in~\eqref{eq:Iperpnu} are to be understood as the averages
of their limits from both sides. This is the result obtained for any
smoothing $\delta_\epsilon$ of the $\delta$-function: the jump of the
derivative is then distributed according to
$\Theta_\epsilon(u)=\int_{-\infty}^u\dd u'\,\delta_\epsilon(u')$, and
$\int\dd u\,\delta_\epsilon(u)\,\Theta_\epsilon(u)=\frac{1}{2}$ independently of
the profile.

The proper-time integrals in~\eqref{eq:gammadyn} become
$\int_0^T\dd\tau_1\int_0^{\tau_1}\dd\tau_2\to\int_0^T\dd u\,(T-u)$. Introducing
$s_1=u$ and $s_2=T-u$, using the symmetry of~\eqref{eq:Iperpnu} under
$s_1\leftrightarrow s_2$, and the Gaussian identity
\begin{equation}
\big[4\pi(s_1+s_2)\big]^{-\frac{d-1}{2}}\,e^{-k_0^2\frac{s_1s_2}{s_1+s_2}}
\;=\; \int\frac{\dd^{d-1}p}{(2\pi)^{d-1}}\;e^{-s_1p^2-s_2(p+k)^2}\;,\qquad
k\equiv(k_0,0,\dots,0)\;,
\end{equation}
where $p$ runs over the $d-1$ momenta conjugate to $x_0$ and to the axial
coordinates, we obtain
\begin{equation}\label{eq:bubble}
\gamma(k_0) \;=\; -\frac{\lambda^2}{2}\sum_\nu\int\frac{\dd^{d-1}p}{(2\pi)^{d-1}}\;
\partial_\rho\partial_{\rho'}\Big[\rho\rho'\,{\mathcal G}_\nu(p^2;\rho,\rho')\,
{\mathcal G}_\nu\big((p+k)^2;\rho',\rho\big)\Big]_{\rho=\rho'=a}\;.
\end{equation}
This is a one-loop polarization diagram built with the radial Green's
functions of Sect.~\ref{sec:arcgreen}, with a vertex
$\lambda\,\partial_\rho(\rho\,\cdot\,)\big|_{\rho=a}$ at each insertion. Its
imaginary part follows from the spectral representation
${\mathcal G}_\nu(E;\rho,\rho')=\int_0^\infty\dd k\,f_{\nu,k}(\rho)f_{\nu,k}(\rho')/(k^2+E)$,
where $f_{\nu,k}$ are the real improper eigenfunctions of $h_\nu$,
$h_\nu f_{\nu,k}=k^2f_{\nu,k}$, normalized as
$\int_0^\infty\dd\rho\,\rho\,f_{\nu,k}(\rho)f_{\nu,k'}(\rho)=\delta(k-k')$ (for
$\lambda\geq0$ the spectrum is purely continuous and nondegenerate). After
the rotation described at the beginning of this section, the total pair
creation probability, $P=2\,\mathrm{Im}\,\Gamma^{(M)}$, reads
\begin{equation}\label{eq:Ptotal}
P \;=\; L^{d-2}\int_0^\infty\frac{\dd\omega}{2\pi}\,\big|\widetilde{R}(\omega)\big|^2
\,{\mathcal P}(\omega)\;,
\end{equation}
where $\widetilde{R}(\omega)\equiv\int\dd t\,e^{\ii\omega t}R(t)$ now denotes
the Fourier transform of the real-time displacement $R(t)$, and
\begin{equation}\label{eq:Pomega}
{\mathcal P}(\omega) \;=\; \frac{\pi\lambda^2}{4}\sum_\nu\int_0^\infty\dd k_1
\int_0^\infty\dd k_2\int\frac{\dd^{d-2}q}{(2\pi)^{d-2}}\;
\frac{D_\nu(k_1,k_2)^2}{\omega_1\,\omega_2}\;\delta(\omega-\omega_1-\omega_2)\;,
\end{equation}
where $\omega_i=\sqrt{k_i^2+q^2}$, $q$ is the axial momentum, and
\begin{equation}\label{eq:Dnu}
D_\nu(k_1,k_2) \;\equiv\; \partial_\rho\big[\rho\,f_{\nu,k_1}(\rho)\,
f_{\nu,k_2}(\rho)\big]_{\rho=a}\;.
\end{equation}
The same result follows from first-order time-dependent perturbation theory
for the emission of a pair of quanta with radial momenta $k_1$, $k_2$ and
opposite axial momenta. Note that both quanta belong to the same angular
sector, since the perturbation does not depend on $\theta$.

\section{The narrow-wedge limit}\label{sec:narrow}

For a small opening angle, $\alpha\ll1$ (that is, $n\gg1$), the wedge is,
near the arc, a slab of local width $\alpha\rho$, and one expects the
angular dependence of the field to be frozen into the lowest mode, the
problem reducing to a lower-dimensional one with a moving $\delta$-mirror.
In this section we show that the reduction to angular sectors is in fact
exact, and that the narrow-wedge limit determines which sectors contribute
and allows for an explicit evaluation of the results of
Sects.~\ref{sec:zero} and~\ref{sec:dyn}. This provides both a physical
picture and an independent check of those results.

\subsection{Angular sectors}\label{sec:sectors}
For Dirichlet walls, we expand
\begin{equation}
\varphi(x) \;=\; \sum_{\nu=1}^{\infty}\chi_\nu(\theta)\,\rho^{-\frac{1}{2}}\,
\psi_\nu(x_\parallel,\rho)\;,\qquad
\chi_\nu(\theta)\,=\,\sqrt{\frac{2}{\alpha}}\,\sin(\mu_\nu\theta)\;,
\end{equation}
where $x_\parallel\equiv(x_0,x_1,\dots,x_{d-2})$. Since $V_\lambda$ does not
depend on $\theta$, also when the arc moves, the action splits exactly into
a sum over sectors, ${\mathcal S}=\sum_\nu{\mathcal S}_\nu$, with
\begin{equation}\label{eq:Snu}
{\mathcal S}_\nu \;=\; \frac{1}{2}\int\dd^{d-1}x_\parallel\int_0^\infty\dd\rho\;
\psi_\nu\Big[-\partial_\parallel^2-\partial_\rho^2
+\frac{\mu_\nu^2-\frac{1}{4}}{\rho^2}
+\lambda\,\delta\big(\rho-a-R(x_0)\big)\Big]\psi_\nu\;.
\end{equation}
The factor $\rho^{-1/2}$ removes the first-order radial derivative and
renders the measure flat; the arc term
$\lambda(a+R)\,\rho^{-1}\psi_\nu^2\big|_{\rho=a+R}$ becomes simply
$\lambda\,\psi_\nu^2(a+R)$. Therefore $\Gamma=\sum_\nu\Gamma_\nu$, to all orders
in $R$ and $\lambda$, where $\Gamma_\nu$ is the effective action of a scalar
field in $d$ dimensions ($d-1$ coordinates $x_\parallel$ and $\rho>0$),
subject to an inverse-square potential, regular at the edge,
$\psi_\nu\sim\rho^{\mu_\nu+\frac{1}{2}}$, and coupled to a $\delta$-mirror at
$\rho=a+R(x_0)$. For $\mu_\nu\gg1$, the potential is a position-dependent
mass, $m_\nu(\rho)\simeq\mu_\nu/\rho=\nu\pi/(\alpha\rho)$: the transverse
momentum of the $\nu$-th mode in a slab of width $\alpha\rho$. The
narrow-wedge limit is therefore the large-order limit of each radial
problem, controlled by the uniform asymptotic expansions of the Bessel
functions~\cite{DLMF}.

For mixed walls the angular orders are $\mu_\nu=n(\nu+\frac{1}{2})$,
$\nu=0,1,\dots$, and everything below applies with this replacement; in
particular, the results for the lowest sector hold with $n\to n/2$, while
consecutive orders are still separated by $n$. For Neumann walls, instead, there is a sector
with $\mu_0=0$, whose inverse-square potential, $-\frac{1}{4\rho^2}$, does not
grow with $n$; that sector is not suppressed in the narrow-wedge limit and
requires a separate treatment.

\subsection{Static energy}\label{sec:narrowstatic}
Integrating~\eqref{eq:Elambda} by parts in $E$ (the boundary terms are
scale-less), the energy of the arc becomes a sum of sector contributions,
\begin{equation}\label{eq:Enu}
{\mathcal E}_\lambda \,=\, \sum_\nu{\mathcal E}^{(\nu)}_\lambda\;,\qquad
{\mathcal E}^{(\nu)}_\lambda \,=\, \frac{1}{2}\int\frac{\dd^{d-1}p}{(2\pi)^{d-1}}\,
\log\Big[1+\lambda a\,I_{\mu_\nu}(|p|a)\,K_{\mu_\nu}(|p|a)\Big]\;,
\end{equation}
where $a\,I_{\mu_\nu}(|p|a)K_{\mu_\nu}(|p|a)$ is the propagator of the sector
evaluated at the arc. The uniform asymptotic expansion~\cite{DLMF} gives
\begin{equation}\label{eq:IKdebye}
a\,I_\mu(|p|a)\,K_\mu(|p|a) \;=\; \frac{1}{2\sqrt{p^2+M^2}}\,
\Big[1+\frac{t^2(1-t^2)(1-5t^2)}{8\mu^2}+O\big(\mu^{-4}\big)\Big]\;,
\end{equation}
where $M=\mu/a$ and $t=M/\sqrt{p^2+M^2}$. Hence, at leading order, ${\mathcal E}^{(\nu)}_\lambda$ is the energy per
unit area of a planar $\delta$-mirror,
\begin{equation}\label{eq:Eflat}
{\mathcal E}_\mathrm{flat}(M) \;=\; \frac{1}{2}\int\frac{\dd^{d-1}p}{(2\pi)^{d-1}}\,
\log\Big[1+\frac{\lambda}{2\sqrt{p^2+M^2}}\Big]\;,
\end{equation}
for a field of mass $M_\nu=\mu_\nu/a=n\nu/a$ in $d$ spacetime dimensions: in
the narrow-wedge limit, each angular sector sees the arc as a planar
$\delta$-mirror, the masses being those of the tower of transverse modes of
the local slab of width $\alpha a$. In $d=3$ the agreement is better than
suggested by~\eqref{eq:IKdebye}. On the one hand, the term of first order
in $\lambda$ coincides exactly with its planar counterpart for every
$\mu$, since the Mellin transform
\begin{equation}
\int_0^\infty\dd x\;x^{s-1}\,I_\mu(x)\,K_\mu(x) \;=\;
\frac{\Gamma\big(\frac{s}{2}\big)\,\Gamma\big(\mu+\frac{s}{2}\big)\,
\Gamma\big(\frac{1-s}{2}\big)}{4\sqrt{\pi}\;\Gamma\big(1+\mu-\frac{s}{2}\big)}
\end{equation}
continued to $s=2$ equals $-\frac{\mu}{2}$, the dimensionally regularized
value of the corresponding planar integral. On the other hand, the
difference between the second-order terms is finite and independent of $a$.
As a consequence, the radial force on the arc in sector $\nu$,
$f_\nu=-\partial_a{\mathcal E}^{(\nu)}_\lambda$, agrees with its planar
counterpart up to relative corrections of order $(\lambda a)^2\mu_\nu^{-4}$:
for $\lambda a=2$, the relative difference is $2.4\times10^{-5}$ for
$\mu_\nu=5$, $1.7\times10^{-6}$ for $\mu_\nu=10$ and $7.2\times10^{-9}$ for
$\mu_\nu=40$. Note, however, that the static energy is not dominated by the
lowest sector, since all of them contribute at the same order; sector
dominance is a property of the low-frequency response, to which we now turn.

\subsection{Pair creation}\label{sec:narrowdyn}
The pair creation probability~\eqref{eq:Pomega} involves, in each sector,
the radial modes of $h_\nu$. For a semitransparent arc they are
\begin{equation}\label{eq:modes}
f_{\nu,k}(\rho) \;=\; A_\nu(k)\times\left\{\begin{array}{ll}
J_\mu(k\rho)\;, & \rho<a\;,\\[4pt]
J_\mu(k\rho)+\beta\,J_\mu(ka)\big[J_\mu(ka)Y_\mu(k\rho)-Y_\mu(ka)J_\mu(k\rho)\big]\;,
& \rho>a\;,
\end{array}\right.
\end{equation}
with $\mu=\mu_\nu$, $\beta=\frac{\pi\lambda a}{2}$, and
$A_\nu(k)=\sqrt{k}\,\big[(1-\beta\,J_\mu Y_\mu)^2+\beta^2J_\mu^4\big]^{-1/2}$, the
Bessel functions in $A_\nu$ being evaluated at $ka$. In the Dirichlet limit,
$\lambda D_\nu\to a\big[f'_{\nu,k_1}(a^+)f'_{\nu,k_2}(a^+)-f'_{\nu,k_1}(a^-)
f'_{\nu,k_2}(a^-)\big]$, and the interior and exterior regions decouple. The
exterior modes form a continuum, with
$|f'_{\nu,k}(a^+)|=2\sqrt{k}/\big(\pi a\,|H_\mu(ka)|\big)$,
$|H_\mu|^2=J_\mu^2+Y_\mu^2$, so that
\begin{equation}\label{eq:Pext}
{\mathcal P}^\mathrm{ext}_\nu(\omega) \;=\; \frac{4}{\pi^3a^2}\int_0^\infty\dd k_1
\int_0^\infty\dd k_2\int\frac{\dd^{d-2}q}{(2\pi)^{d-2}}\;
\frac{k_1k_2\;\delta(\omega-\omega_1-\omega_2)}{\omega_1\,\omega_2\,
|H_\mu(k_1a)|^2\,|H_\mu(k_2a)|^2}\;,
\end{equation}
while the interior modes are discrete, $k_m=j_{\mu,m}/a$, with
$|f'_{\nu,m}(a^-)|=\sqrt{2}\,k_m/a$, and yield
\begin{equation}\label{eq:Pint}
{\mathcal P}^\mathrm{int}_\nu(\omega) \;=\; \frac{\pi}{a^2}\sum_{m,m'}
\int\frac{\dd^{d-2}q}{(2\pi)^{d-2}}\;\frac{k_m^2\,k_{m'}^2}{\omega_m\,\omega_{m'}}\;
\delta(\omega-\omega_m-\omega_{m'})\;,\qquad \omega_m=\sqrt{k_m^2+q^2}\;.
\end{equation}
For $d=2$ the interior contributes lines at $\omega=(j_{\mu,m}+j_{\mu,m'})/a$,
the parametric resonances of the cavity; for $d>2$ the axial momentum turns
each line into a band starting at that frequency.

Let us now take $n\to\infty$. For $ka>\mu$, and away from the transition
region $|ka-\mu|\lesssim\mu^{1/3}$,
$|H_\mu(ka)|^2=\frac{2}{\pi}\big(k^2a^2-\mu^2\big)^{-1/2}\big[1+O(\mu^{-2})\big]$,
while for $ka<\mu$ it grows exponentially with $\mu$~\cite{DLMF}. Introducing
the normal momentum $\bar{k}=\sqrt{k^2-M_\nu^2}$, the exterior spectrum tends
to
\begin{equation}\label{eq:Pflat}
{\mathcal P}_\mathrm{flat}(\omega;M) \;=\; \frac{1}{\pi}\int_0^\infty\dd\bar{k}_1
\int_0^\infty\dd\bar{k}_2\int\frac{\dd^{d-2}q}{(2\pi)^{d-2}}\;
\frac{\bar{k}_1^2\,\bar{k}_2^2}{\omega_1\,\omega_2}\;\delta(\omega-\omega_1-\omega_2)\;,
\end{equation}
with $\omega_i=\sqrt{\bar{k}_i^2+M^2+q^2}$. This is the spectrum of a planar
Dirichlet mirror oscillating along its normal and
radiating to one side, for a field of mass $M=M_\nu$ in $d$ spacetime
dimensions; for $M=0$ and $d=2$ it reduces to $\omega^3/(6\pi)$. Thus, in the
narrow-wedge limit, the exterior region behaves as a flaring waveguide whose
lowest mode has the cutoff $n/a=\pi/(\alpha a)$, and pairs in sector
$\nu$ are created for $\omega>2\mu_\nu/a$.

The leading correction comes from the transition regions. For $d=2$, writing
$x=\mu+\zeta\,(\mu/2)^{1/3}$, one has $|H_\mu(x)|^{-2}\simeq(\mu/2)^{2/3}
\big[\mathrm{Ai}^2(-\zeta)+\mathrm{Bi}^2(-\zeta)\big]^{-1}$, and the excess of
$|H_\mu(x)|^{-2}$ over its asymptotic form
$\frac{\pi}{2}\sqrt{x^2-\mu^2}\,\theta(x-\mu)$ integrates to
$\frac{\mu}{2}C_A$, with
\begin{equation}
C_A \;=\; \int_{-\infty}^{\infty}\dd\zeta\,\Big\{\big[\mathrm{Ai}^2(-\zeta)
+\mathrm{Bi}^2(-\zeta)\big]^{-1}-\pi\sqrt{\zeta}\;\theta(\zeta)\Big\}
\;=\;\frac{\pi^2}{4}\;,
\end{equation}
the last equality holding within our numerical accuracy (ten significant
digits).
Since each of the two thresholds contributes, we find, for the lowest sector,
\begin{equation}\label{eq:Pcorr}
{\mathcal P}^\mathrm{ext}_1(\omega) \;=\; {\mathcal P}_\mathrm{flat}(\omega;n/a)\,
\Big[1+\frac{c(\Omega)}{n}+O\big(n^{-2}\big)\Big]\;,\qquad
c(\Omega)\,=\,\frac{\pi}{2}\,\frac{\sqrt{(\Omega-1)^2-1}}{F(\Omega)}\;,
\end{equation}
where $\Omega\equiv\omega a/n$,
$F(\Omega)=\int_1^{\Omega-1}\dd x\,\sqrt{x^2-1}\,\sqrt{(\Omega-x)^2-1}$, and
${\mathcal P}_\mathrm{flat}(\omega;n/a)=\frac{n^3}{\pi a^3}F(\Omega)$. At $\Omega=3$,
$c=2.785$, while a direct numerical evaluation of~\eqref{eq:Pext} gives
$n\big[{\mathcal P}^\mathrm{ext}_1/{\mathcal P}_\mathrm{flat}-1\big]=2.669$, $2.724$,
$2.754$, $2.770$ and $2.777$ for $n=10$, $20$, $40$, $80$ and $160$.

\begin{figure}[t]
\centering
\includegraphics[width=\textwidth]{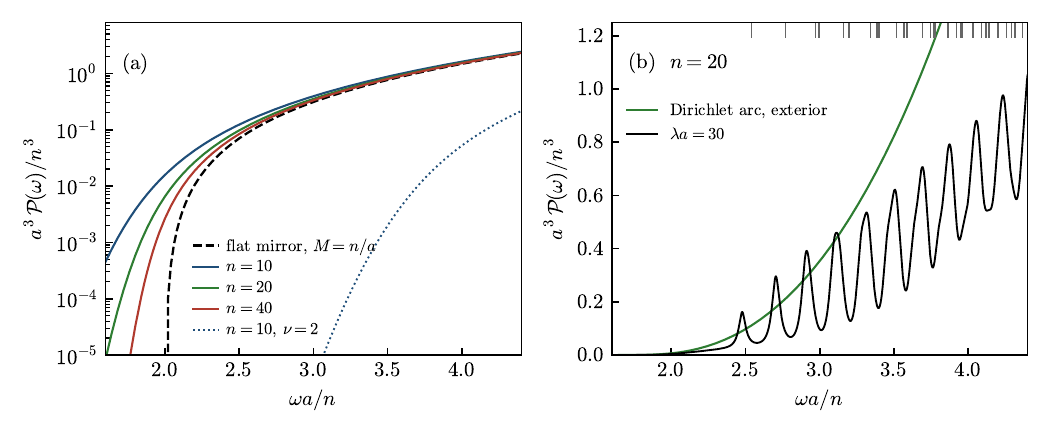}
\caption{Pair creation spectrum ${\mathcal P}(\omega)$ of the lowest angular
sector, for $d=2$, in units of $n^3/a^3$, as a function of $\omega a/n$.
(a) Dirichlet arc, exterior region, Eq.~\eqref{eq:Pext}, for $n=10$, $20$, $40$
(solid lines), compared with the planar massive mirror~\eqref{eq:Pflat} with
$M=n/a$ (dashed line); the dotted line is the contribution of the sector
$\nu=2$ for $n=10$. (b) Semitransparent arc with $\lambda a=30$ and $n=20$,
from~\eqref{eq:Pomega} and~\eqref{eq:modes}, compared with the exterior
contribution for a Dirichlet arc; the ticks at the top mark the interior
lines of the Dirichlet arc, at $\omega a/n=(j_{n,m}+j_{n,m'})/n$.}
\label{fig:narrow}
\end{figure}

Since pairs are created within a single sector, the lowest one determines
the probability in the window $2n/a\lesssim\omega\lesssim4n/a$; at
$\Omega=3.5$, the contribution of $\nu=2$ relative to that of $\nu=1$ is
$2.3\times10^{-3}$, $5.4\times10^{-5}$ and $3.9\times10^{-8}$ for $n=10$, $20$
and $40$. The edges of the window are smeared by tunneling through the
centrifugal barrier, over frequency intervals of order $n^{1/3}/a$: below
$\omega=2n/a$ the probability does not vanish, but it decreases
exponentially with $n$, see Fig.~\ref{fig:narrow}(a).

The interior region is, instead, a tapered cavity: the radial modes of
sector $\nu$ are confined between the turning point $\rho_\mathrm{t}=\mu_\nu/k$
and the arc. Linearizing the potential near the arc,
$(\mu^2-\frac{1}{4})/\rho^2\simeq M^2-2M^2(\rho-a)/a$, the problem becomes an
Airy cavity (a quantum bouncer) closed by the mirror, whose eigenvalues,
$k_m^2a^2\simeq n^2+2^{\frac{2}{3}}|a_m|\,n^{\frac{4}{3}}$, with $a_m$ the
zeros of the Airy function, reproduce the large-order expansion
$j_{n,m}\simeq n+2^{-\frac{1}{3}}|a_m|\,n^{\frac{1}{3}}$~\cite{DLMF}. For a
Dirichlet arc, the interior lines~\eqref{eq:Pint} start at
$2j_{n,1}/a\simeq(2n+3.71\,n^{1/3})/a$. For a semitransparent arc the
interior modes become resonances, slightly below the Dirichlet values, and
the spectrum shows peaks near $(j_{n,m}+j_{n,m'})/a$ on top of the exterior
background, see Fig.~\ref{fig:narrow}(b). Therefore, a narrow wedge closed
by a semitransparent arc is not equivalent to a free planar $\delta$-mirror
for a massive field: on its inner side, the mirror faces a soft reflecting
wall, created by the angular quantization.

\section{The electromagnetic field}\label{sec:em}

\subsection{Static configurations}\label{sec:emstatic}

We now consider the electromagnetic field inside the wedge, with Euclidean 
action ${\mathcal S}(A) = \frac{1}{4}\int \dd^4x\,F_{\mu\nu}F_{\mu\nu}$, 
$F_{\mu\nu} = \partial_\mu A_\nu - \partial_\nu A_\mu$. Each wall is either a 
perfect electric conductor (PEC), $N_\mu\widetilde{F}_{\mu\nu}=0$ (vanishing 
tangential $\vec{E}$ and normal $\vec{B}$), or a perfect magnetic conductor 
(PMC), $N_\mu F_{\mu\nu}=0$ (vanishing tangential $\vec{B}$ and normal 
$\vec{E}$), $N_\mu$ denoting the unit normal to the wall. Since the walls are 
invariant under translations and rotations in the plane $(x_0,x_1)$, we use 
indices $i,j\in\{0,1\}$ for that plane and $r,s\in\{2,3\}$ for the 
plane orthogonal to the edge, so that $\partial^2 = \partial_\parallel^2 + 
\nabla_\perp^2$. A two-dimensional Helmholtz decomposition in each plane,
\begin{equation}
A_r \;=\; \partial_r\chi+\epsilon_{rs}\,\partial_s\phi\;,\qquad
A_i \;=\; \partial_i\eta+\epsilon_{ij}\,\partial_j\sigma\;,
\end{equation}
yields
\begin{equation}
F_{23} \,=\, -\nabla_\perp^2\phi\;,\qquad 
F_{01} \,=\, -\partial_\parallel^2\sigma\;,\qquad 
F_{ir} \,=\, \partial_i\partial_r u
+\epsilon_{rs}\,\partial_i\partial_s\phi
-\epsilon_{ij}\,\partial_j\partial_r\sigma\;,
\end{equation}
with $u\equiv\chi-\eta$. Under a gauge transformation, $\chi$ and $\eta$ 
shift by the same function, so that $\phi$, $\sigma$ and $u$ are gauge 
invariant. Since the component of $\vec{B}$ along the edge depends only on 
$\phi$, and that of $\vec{E}$ only on $\sigma$, $\phi$ generates the TE fields 
and $\sigma$ the TM fields. A PEC wall requires $\sigma=0$, $\partial_N\phi=0$, 
$u=0$, while a PMC wall requires $\phi=0$, $\partial_N\sigma=0$, $\partial_N 
u=0$. With these conditions, all cross terms in the action vanish after 
integration by parts, and
\begin{equation}
{\mathcal S}(A) \;=\; \frac{1}{2}\int\phi\,(-\nabla_\perp^2)(-\partial^2)\,\phi
+\frac{1}{2}\int\sigma\,(-\partial_\parallel^2)(-\partial^2)\,\sigma
+\frac{1}{2}\int u\,(-\partial_\parallel^2)(-\nabla_\perp^2)\,u\;.
\end{equation}
 The field $u$ does not propagate, and a redefinition of the fields, $\Phi=(-\nabla_\perp^2)^{1/2}\phi$ and $\Sigma=(-\partial_\parallel^2)^{1/2}\sigma$, gives the action of two massless scalar fields with unaltered boundary conditions, $\frac{1}{2}\int(\partial_\mu\Phi\,\partial_\mu\Phi+\partial_\mu\Sigma\,\partial_\mu\Sigma)$. The lower-dimensional determinants produced 
by the Gaussian integrals cancel against the Jacobian of the change of 
variables $A\to(\chi,\phi,\eta,\sigma)$, and the gauge volume factors out, 
so that
\begin{equation}\label{eq:EMfact}
{\mathcal Z}_\mathrm{EM} \;=\; \mathrm{Det}\big(-\partial^2\big|_\mathrm{TE}\big)^{-\frac{1}{2}}
\,\mathrm{Det}\big(-\partial^2\big|_\mathrm{TM}\big)^{-\frac{1}{2}}\;,\qquad 
\Gamma_\mathrm{EM} \;=\; \Gamma_\mathrm{TE}+\Gamma_\mathrm{TM}\;,
\end{equation}
where each term is the effective action of a massless scalar with the 
conditions inherited by $\phi$ or $\sigma$, which may be evaluated 
with~\eqref{eq:cale}. This is the same decomposition obtained in Ref.~\cite{NLS} for a semicircular cylinder boundary. For PEC walls, TE~$\to$~Neumann and TM~$\to$~Dirichlet; 
for PMC walls the roles are exchanged; for PEC at $\theta=0$ and PMC at 
$\theta=\alpha$, the TM scalar is Dirichlet-Neumann and the TE scalar
Neumann-Dirichlet, which has the same spectrum. In $d+1$ dimensions, the same
construction gives one TE and $d-2$ TM scalars, i.e., the $d-1$ physical
polarizations; since the scalar energies of the open wedge have no pole at
$d=3$, this does not modify the results below. This is no longer true for the
closed sectors considered at the end of this section, whose scalar energies
may have a pole at $d=3$: there, the finite part depends on how the
polarizations are counted away from $d=3$.

From Sect.~\ref{sec:zero}, the radial energy densities are
\begin{align}
\varepsilon_\mathrm{PEC}(\rho) \,=\, \varepsilon_\mathrm{PMC}(\rho) &\,=\, 
2\,\varepsilon_\mathrm{D}(\rho) \,=\, 
-\frac{(n^2-1)(n^2+11)}{720\,\pi\,n}\;\frac{1}{\rho^3}\;,\label{eq:EMpure}\\
\varepsilon_\mathrm{PEC/PMC}(\rho) &\,=\, 2\,\varepsilon_\mathrm{DN}(\rho) \,=\, 
\frac{7n^4+40n^2+88}{5760\,\pi\,n}\;\frac{1}{\rho^3}\;.\label{eq:EMmixed}
\end{align}
For the electromagnetic field, the energy-momentum tensor is the Maxwell 
tensor, which is fixed by gauge invariance and admits no improvement of the 
$\xi$ type. Indeed, \eqref{eq:EMpure} is exactly the integral of the
Deutsch-Candelas energy density~\cite{DC}
\begin{equation}
\langle T^{00}\rangle_\mathrm{PEC} \;=\; -\frac{(n^2-1)(n^2+11)}{720\,\pi^2\rho^4}\;.
\end{equation}
At the local level, this density equals the sum of the Dirichlet and Neumann 
scalar densities with $\xi=1/4$: the wall-divergent terms cancel between the 
two scalars, while the $\theta$-independent ones add up. For general $\xi$ the 
sum is $-\frac{(n^2-1)(n^2+120\xi-19)}{720\pi^2\rho^4}$, which reproduces the 
electromagnetic result only for $\xi=1/4$; in particular, the electromagnetic 
field in the wedge is not equivalent to two conformally coupled scalars. In the 
same way, for the PEC/PMC wedge the sum of the two mixed $\xi=1/4$ densities is
\begin{equation}
\langle T^{00}\rangle_\mathrm{PEC/PMC} \;=\; \frac{7n^4+40n^2+88}{5760\,\pi^2\rho^4}\;,
\end{equation}
again independent of $\theta$ and finite on the walls. For a narrow wedge 
($n\to\infty$), the local separation between the walls is $\rho\alpha$, and 
these densities reduce to $-\frac{\pi^2}{720(\rho\alpha)^4}$ and 
$+\frac{7\pi^2}{5760(\rho\alpha)^4}$, the energy densities between two 
perfectly conducting plates~\cite{Casimir,BrownMaclay} and between a 
perfectly conducting and an infinitely permeable plate~\cite{Boyer}, 
respectively (in the latter case, the density follows from the force 
obtained in Ref.~\cite{Boyer}, since the energy density between parallel 
plates is uniform).

Finally, a combination of two Dirichlet scalars corresponds to the 
soft-and-hard surface (SHS) condition with axis along the 
edge~\cite{Kildal}, which imposes $E_z=H_z=0$ on the walls, $z$ denoting the 
coordinate along the edge. For the open wedge, its energy coincides with that 
of the PEC wedge, since ${\mathcal E}_\mathrm{N}={\mathcal E}_\mathrm{D}$.

The arc is parallel to the edge, so the decomposition~\eqref{eq:EMfact} still 
holds when it is present, now with a closed transverse domain. A PEC arc 
imposes Dirichlet conditions on the TM scalar and Neumann conditions on the TE 
scalar, while a PMC arc does the opposite. Thus, any PEC or PMC sector 
requires a scalar with a Neumann condition on the arc, whose energy was 
obtained in Sect.~\ref{sec:arcN}. 
The electromagnetic energy of a closed sector is then the sum of the two scalar 
energies. For instance, for PEC walls and arc,
\begin{equation}
{\mathcal E}_\mathrm{PEC} \;=\; {\mathcal E}^\mathrm{arc}_\mathrm{D}\Big|_{\mu_\nu=n\nu,\;
\nu\geq1} + {\mathcal E}^\mathrm{arc}_\mathrm{N}\Big|_{\mu_\nu=n\nu,\;\nu\geq0}\;,
\end{equation}
while for PMC walls and arc the conditions on the arc are exchanged between 
the two scalars. The pole counting of Sect.~\ref{sec:arcgreen} applies to
both terms: the TM contribution (Dirichlet walls and arc, $\zeta(0)=-\frac{1}{2}$
orders) and the TE one (Neumann walls and arc, $1+\zeta(0)=\frac{1}{2}$ orders)
have poles at $d=3$ with residues of the same sign, which add up for every
$\alpha$. Hence the energy of a PEC sector contains an $\alpha$-independent
logarithmic divergence associated with the corners, in agreement with the
divergence found for the semicircular cylinder ($n=1$) in Ref.~\cite{NLS}. For
the full cylinder, instead, the zeta-regularized number of angular orders is
$1+2\zeta(0)=0$, and the TE and TM contributions are separately finite; their
sum reproduces the known value $-0.0135613/a^2$ for the perfectly conducting
cylinder (Eq.~(45) of Ref.~\cite{GosdzinskyRomeo}).

\subsection{Moving arc}\label{sec:emdyn}

The decomposition~\eqref{eq:EMfact} relies on the invariance of the
boundaries under translations and rotations in the $(x_0,x_1)$ plane, which
a moving arc breaks. On the surface $\rho=a+R(x_0)$ the unit normal
$N_\mu$ acquires a component along $x_0$, and the condition
$N_\mu\widetilde{F}_{\mu\nu}=0$ for a moving PEC arc couples the TE and TM
scalars at first order in $\dot{R}$, for modes with nonvanishing axial
momentum and angular order (in Minkowski language, the condition becomes
$\hat{\rho}\times(\vec{E}+\dot{R}\,\hat{\rho}\times\vec{B})=0$). The
dynamical problem does not reduce, therefore, to two independent scalars
with moving Dirichlet and Neumann conditions, and the mixed TE-TM
contributions must be included at second order in $R$. Regarding the
narrow-wedge limit, the TM scalar has Dirichlet walls, and the analysis of
Sect.~\ref{sec:narrow} applies to it; the TE scalar, on the other hand, has
Neumann walls and thus a sector with $\mu_0=0$, which is not suppressed.
Consequently, a narrow wedge with PEC walls behaves as a parallel-plate
waveguide whose lowest mode has no cutoff. A complete account of the
electromagnetic pair creation is left for future work.

\section{Conclusions}\label{sec:conclusions}

In this paper we have studied a real scalar field confined to a wedge,
optionally closed by a semitransparent circular arc, using the worldline
formulation. Our results can be summarized as follows.

At the static level, the zeroth-order effective action is determined by the
propagator of a non-relativistic particle in the wedge, with the boundary
conditions of the corresponding scalar problem. This yields the energies of
wedges with Dirichlet, Neumann or mixed walls, and, for the arc, the closed
expression~\eqref{eq:Elambda}, exact in the strength $\lambda$ of the
$\delta$-potential, whose strong-coupling limit gives the energy of the
Dirichlet arc, Eq.~\eqref{eq:EDarc}; the energy of the Neumann arc,
Eq.~\eqref{eq:ENarc}, follows from the corresponding propagator. For
perfectly reflecting arcs, the only divergence that remains in dimensional
regularization is a pole at $d=3$ associated with the corners where the arc
meets the walls; it is
independent of $\alpha$, so that the torque on the walls is finite, it is
absent for mixed walls, and it cancels for the full cylinder, for which the
known scalar and electromagnetic energies of Ref.~\cite{GosdzinskyRomeo} are
reproduced.

We have also shown that the worldline integrand admits a local
interpretation: point by point inside the wedge, it equals the energy
density of the scalar field with curvature coupling $\xi=\frac{1}{4}$. Since
the proof uses only the spectral representation of the integrand and the
form of the improved energy-momentum tensor, it holds for any static
potential. This identification turns the known arbitrary-$\xi$ results of
Refs.~\cite{DC,RS} into exact evaluations of the worldline integrand,
explains the $\xi$-dependence of the radial energy density through a
surface term at the edge, and accounts for the vanishing, at
$\xi=\frac{1}{4}$, of the leading divergence at the corners found in
Ref.~\cite{RS}. For the electromagnetic field, the Deutsch-Candelas density
is the sum of the Dirichlet and Neumann densities with $\xi=\frac{1}{4}$.

At the dynamical level, we have obtained the first- and second-order terms
in the expansion of the effective action in the radial displacement of the
arc. The first one is proportional to $\widetilde{R}(0)$ and vanishes when
$a$ is the mean radius; the second one is determined by the form factor
$\gamma(k_0)$, which we have reduced to a one-loop expression in terms of
the radial Green's functions of each angular sector,
Eq.~\eqref{eq:bubble}. Its imaginary part gives the pair creation
probability in terms of the radial modes, Eq.~\eqref{eq:Pomega}, both quanta
of a pair belonging to the same angular sector. In the narrow-wedge limit,
each sector becomes a field with a position-dependent mass $\mu_\nu/\rho$
coupled to a moving $\delta$-mirror. The static energy of each sector
approaches that of a planar $\delta$-mirror for a field of mass $\mu_\nu/a$,
and, for a Dirichlet arc, the exterior pair creation spectrum approaches
that of a planar massive mirror, with corrections of order $1/n$ that, for
$d=2$, we have obtained in closed form (the value of the constant $C_A$
being established numerically); the interior region behaves as a tapered
cavity, which, for a semitransparent arc, produces resonances in the
spectrum.

Several directions remain open. The explicit evaluation of~\eqref{eq:Pomega}
for arbitrary $\alpha$ and $\lambda$ in $d=3$, and its electromagnetic
extension to a moving conducting arc, which requires the inclusion of the
TE-TM mixing induced by the motion, are the natural continuation of
Sects.~\ref{sec:dyn} and~\ref{sec:em}. A $\delta'$-coupling on the arc, or
a suitable combination of $\delta$ and $\delta'$ terms, would give an
independent route to the Neumann arc of Sect.~\ref{sec:arcN}. The relation
between the $\alpha$-derivative of the energy obtained here, which is the
global energy of the $\xi=\frac{1}{4}$ density regularized without any
splitting in time, and the torque obtained from the stress
tensor~\cite{FMT,Dowker_torque,Milton_annular1,Milton_annular2}, deserves a
separate study. The radial force on the arc follows from~\eqref{eq:Elambda}
by differentiation with respect to $a$; for a Dirichlet arc it is finite
for mixed walls, and could be compared with the results of
Refs.~\cite{SaharianTarloyan,MWK}. Finally, the worldline integrand
generalizes directly to nonzero temperature and to massive fields, where
closed forms are no longer available and worldline Monte Carlo
methods~\cite{Gies_WL} become the natural tool, also for arbitrary $\lambda$
and $\alpha$.

\section*{Acknowledgments}
This work was supported by CONICET and UNCuyo.

\end{document}